\documentclass[%
 amsmath,amssymb,
 aps, twocolumn]{revtex4-1}
\usepackage{CJK}
\usepackage{graphicx}% Include figure files
\usepackage{dcolumn}% Align table columns on decimal point
\usepackage{bm}% bold math

\usepackage{scalerel} 
\usepackage{titlesec}
\usepackage{float}
\usepackage{amssymb}
\usepackage{standalone}
\usepackage{siunitx}

\newcommand{\RomanNumeralCaps}[1]
    {\MakeUppercase{\romannumeral #1}}

\titlespacing*{\section}
{0pt}{2.5ex plus 1ex minus .2ex}{2.3ex plus .2ex}
\titlespacing*{\subsection}
{0pt}{2.5ex plus 1ex minus .2ex}{2.3ex plus .2ex}

\def\app#1#2{%
  \mathrel{%
    \setbox0=\hbox{$#1\sim$}%
    \setbox2=\hbox{%
      \rlap{\hbox{$#1\propto$}}%
      \lower1.1\ht0\box0%
    }%
    \raise0.25\ht2\box2%
  }%
}

\begin{document}

%\preprint{AIP/123-QED}
\begin{CJK*}{GB}{}
\title[Modelling Enclosures for Large-Scale Superconducting Quantum Circuits]{Modelling Enclosures for Large-Scale Superconducting Quantum Circuits}% 
\author{P. A. Spring}
\author{T. Tsunoda}
\author{B. Vlastakis}
\author{P. J. Leek}
\affiliation{Clarendon Laboratory, Department of Physics, University of Oxford, Oxford OX1 3PU, United Kingdom}

\date{\today}
\begin{abstract}
Superconducting quantum circuits are typically housed in conducting enclosures in order to control their electromagnetic environment. As devices grow in physical size, the electromagnetic modes of the enclosure come down in frequency and can introduce unwanted long-range cross-talk between distant elements of the enclosed circuit. Incorporating arrays of inductive shunts such as through-substrate vias or machined pillars can suppress these effects by raising these mode frequencies. Here, we derive simple, accurate models for the modes of enclosures that incorporate such inductive-shunt arrays. We use these models to predict that cavity-mediated inter-qubit couplings and drive-line cross-talk are exponentially suppressed with distance for arbitrarily large quantum circuits housed in such enclosures, indicating the promise of this approach for quantum computing. We find good agreement with a
finite-element simulation of an example device containing more than 400 qubits.
\end{abstract}
\maketitle
\end{CJK*}

%%%% start of main text

\section{INTRODUCTION}
%%%
Superconducting circuits are a promising platform for quantum computing. Their success originates partly from the intrinsically strong coupling of superconducting qubits to electromagnetic fields, which itself derives from their macroscopic size. This strong coupling facilitates fast quantum logic gates \cite{sheldon2016characterizing, barends2019diabatic}, readout \cite{heinsoo2018rapid} and reset of qubits \cite{reed2010fast}. However, it also makes superconducting qubits prone to couple to spurious electromagnetic (EM) modes in their environment. This can cause deleterious effects such as radiative energy relaxation \cite{houck2008controlling}, coherent leakage of the qubit state \cite{mcconkey2018mitigating}, and mediation of undesired inter-qubit couplings \cite{filipp2011multimode}. It is therefore important to engineer their environment such that couplings to spurious EM modes are suppressed. 
A powerful solution is to house quantum circuits in a cavity that has a fundamental (lowest) mode frequency well above qubit frequencies \cite{ paik2011observation}.
\\
As superconducting circuits grow in size, the fundamental frequency of a simple contigious cavity enclosure must come down. This can be mitigated to some degree by designing vacuum regions into the cavity that reduce the field energy stored in the high dielectric substrate on which the circuit is fabricated \cite{bronn2018high, wenner2011wirebond}, but this approach is not scalable. An approach that is known to be scalable is to inductively shunt the cavity with an array of through-substrate vias (TSVs) \cite{gambetta2017building,vahidpour2017superconducting, yost2019solid}, which must be bonded in some way to both sides of the cavity. The scalability of this approach is related to the physics of metallic photonic crystals \cite{nicorovici1995photonic, smith1994experimental}, which predicts a scale-independent cut-off frequency for this type of periodic metal structure, below which it cannot sustain modes \cite{nicorovici1995photonic}.
\\
This scale-independent cut-off frequency means inductively-shunted cavities can provide a clean EM environment to superconducting circuits at arbitrary size. When engineering the layout of the shunt array, it is then relevant to ask how the cavity mode frequencies depend on the shunt spacing and size, and how cavity-mediated inter-qubit couplings and drive-line cross-talk is affected.
\\
The purpose of this work is to address these questions, by constructing accurate, closed-form, physically intuitive models for the modes of a cavity that is inductively shunted by a square array of cylinders; and to then use these results to predict cavity-mediated inter-qubit couplings and drive-line cross-talk inside such a cavity.
\\
The paper is arranged as follows: in section \RomanNumeralCaps{2} we develop a plasma model and a circuit model for cavities with periodic inductive shunt arrays. In section \RomanNumeralCaps{3} we use these models to predict inter-qubit coupling and drive-line cross-talk for superconducting qubits inside such cavities. In section \RomanNumeralCaps{4} we test the predictions against a finite-element (FE) simulation of a simplified device containing a $21\times 21$ grid of qubits, representing a device in the NISQ regime \cite{preskill2018quantum}.
%%%
\section{\label{sec:models} MODELS FOR THE PERIODICALLY INDUCTIVELY SHUNTED CAVITY}
\begin{figure}
  \centering
    \includegraphics[width=0.48\textwidth]{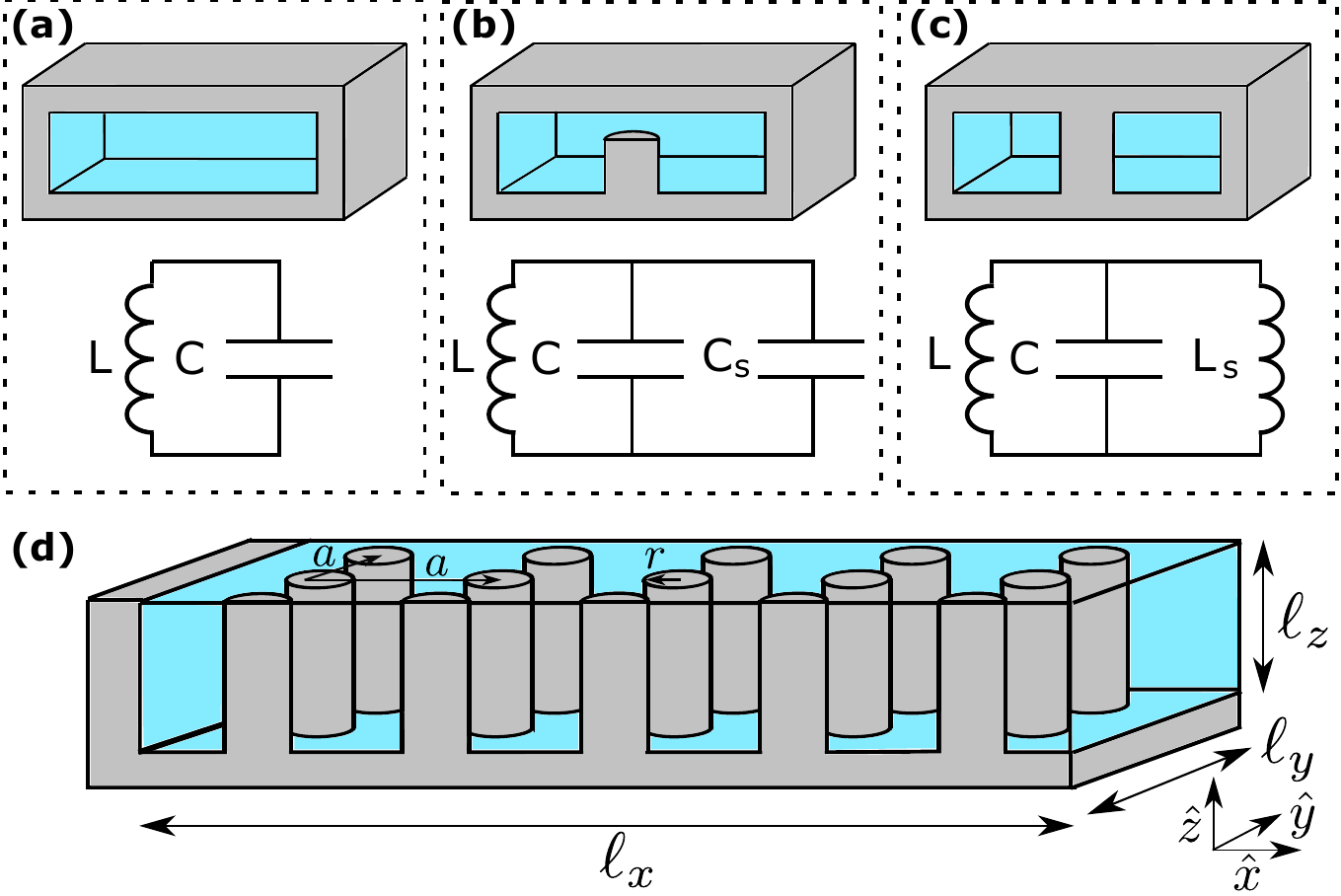}
  \setlength{\belowcaptionskip}{-10pt}
  \vspace{-10pt}
  \caption{\textbf{(a)} - \textbf{(c)} Cross-sections of rectangular cavities along with circuit representations of their fundamental mode. \textbf{(d)} A cavity filled with a dielectric and inductively shunted by a square array of conducting cylinders, spacing $a$ and radius $r$.}
  \label{fig:circuitmodelvias}
\end{figure}

%%%
In fig.~\ref{fig:circuitmodelvias}, we illustrate how the fundamental frequency of a simple rectangular cavity (fig.~1(a)) is altered by the presence of a protruding pillar. An unshorted pillar behaves as a capacitive shunt (fig.~1(b)), decreasing the the cavity mode frequency, whereas a shorted pillar behaves as an inductive shunt (fig.~1(c)), increasing the cavity mode frequency.
\\
We will consider the extension of the single inductive shunt to an inductive shunt array, as shown in fig.~\ref{fig:circuitmodelvias}(d). The array is formed of cylinders radius $r$ and equal spacing $a$ in $\hat{\bm{x}}$ and $\hat{\bm{y}}$. The cavity is perfectly conducting, has dimensions $\ell_x,\ell_y,\ell_z$, and is filled with a uniform material with dielectric permittivity $\epsilon_r$. Multiple dielectric layers stacked along the $\hat{\bm{z}}$ axis can be modelled simply by an adjustment to $\epsilon_r$ (See Appendix A).
\\
In the absence of inductive shunts, the mode frequencies of this cavity are those of a rectangular cavity, given by
\begin{equation}
    f_{nml}= \frac{1}{2\sqrt{\epsilon \mu}}\sqrt{\frac{n^2}{\ell_x^2}+\frac{m^2}{\ell_y^2}+\frac{l^2}{\ell_z^2}}
\label{eq:reccav}
\end{equation}
where $n,m,l$ take integer values, and physical solutions permit only one of $n,m,l$ to be zero. In superconducting quantum devices, the circuit substrate thickness does not scale with the circuit size and so we consider the case $\ell_z \ll \ell_x,\ell_y$. The low-frequency spectrum then consists only of $l=0$ modes, which we will focus on for the remainder of this article.
%%%
\subsection{\label{sec:primitive}Boundary Model}
%%%
As an illustrative model, consider replacing the array of inductive shunts with a grid of thin conducting walls, with spacing $a$. This results in a fundamental frequency
\begin{eqnarray}
f_{a}=1/(a\sqrt{2\epsilon_0 \epsilon_r \mu_0})
\label{eq:PrimativeFundamental}
\end{eqnarray}
This primitive model predicts the existence of a cut-off frequency independent of the total enclosure size, but fails to take into account the shunt radius $r$, which is clearly an oversimplification. In the limit $r \rightarrow 0$ where the shunts disappear, we should instead recover eq. \ref{eq:reccav}.
%%%
\subsection{\label{sec:plasma}Plasma Model $\boldsymbol{(r/a<0.1)}$}
\begin{figure}
  \centering
    \includegraphics[width=0.48\textwidth]{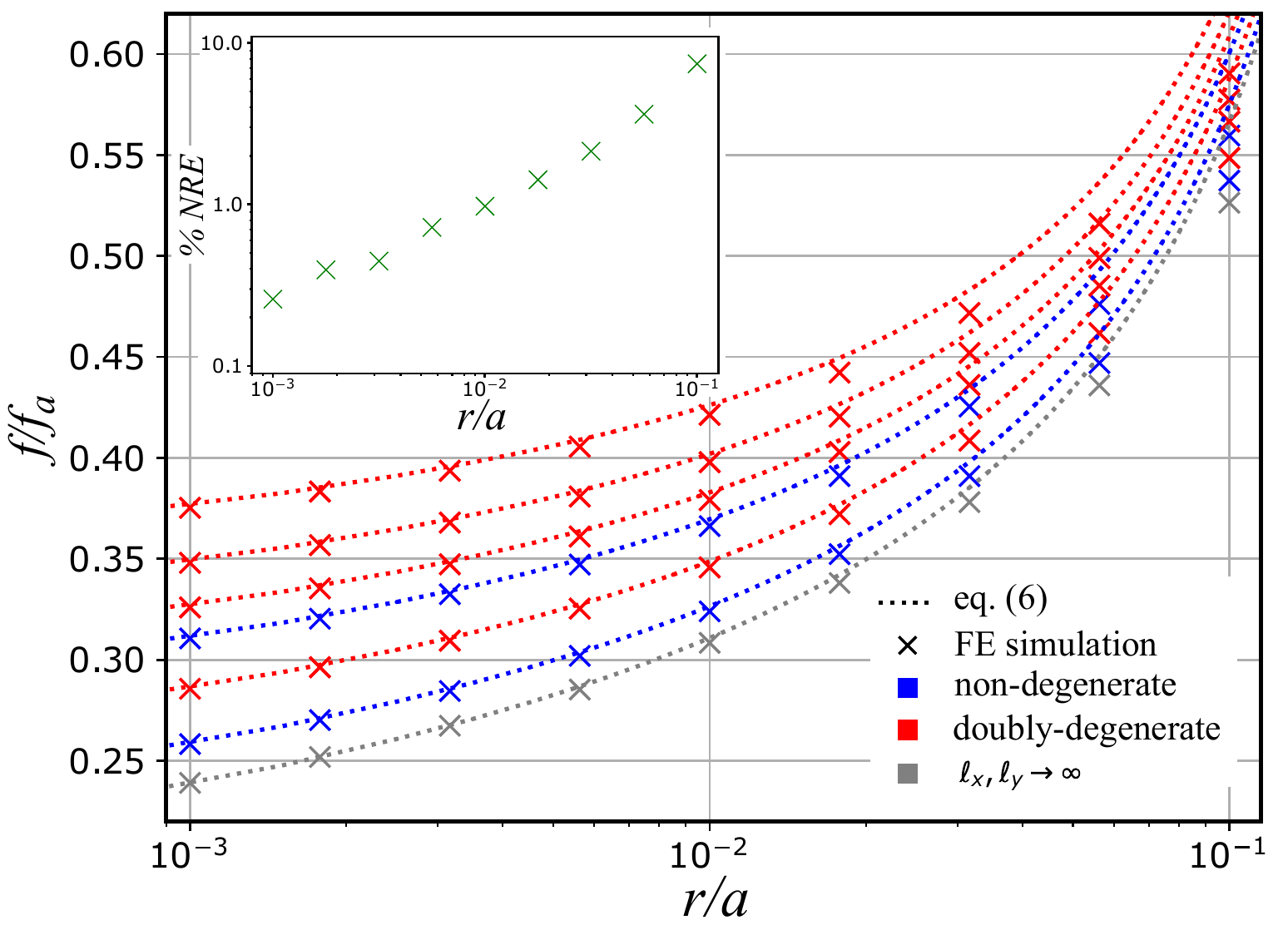}
  \setlength{\belowcaptionskip}{-10pt}
  \vspace{-15pt}
  \caption{Lowest 10 modes for a cavity containing a containing an inductive shunt array, with $\ell_x,\ell_y =10 a$. The mode frequencies and degeneracies are accurately predicted by the plasma model.  Also included in grey are results for the fundamental mode when $\ell_x,\ell_y \rightarrow \infty$, with simulation values found using a method described in Ref. \cite{remski2000analysis}. Inset shows the normalized relative error (NRE) between simulation and eq. (\ref{eq:CavFreq2}) $\Sigma_{i=1}^{10}|f_{i_{\text{FE}}}-f_{i_{\text{eq}}}| /10f_{i_\text{FE}}$.}
  \label{fig:viaLow}
\end{figure}

%%%
The behaviour of an array of thin, infinitely long conducting cylinders (oriented along $\bf{\hat{z}}$) has been studied as a meta-material, and has been shown to behave like an anisotropic plasma in the limit $r \ll a \ll \lambda$, with the associated frequency-dependent permittivity \cite{pendry1996extremely}
\begin{eqnarray}
\bm{\epsilon_p}(f) = (1-(\frac{f_p}{f})^2) \hat{\bm{z}},
\label{eq:PlasmaPerm}
\end{eqnarray}
valid for EM waves propagating in the x-y plane \cite{belov2003strong}. The plasma frequency $ f_p $ is accurately predicted by a simple function of the cylinder radius and spacing \cite{belov2002dispersion,krynkin2009approximations}
\begin{gather}
f_p= \frac{f_a}{\sqrt{\pi}(\ln (\frac{a}{r})-\Pi)^{0.5}}
\label{eq:BelovsFreq}
\\
\Pi = \ln(2\pi)-\pi/6-\Sigma_{n=1}^\infty (\coth{(n\pi)}-1)/n \approx 1.31
\nonumber
\end{gather}
We wish to apply these equations to the $l=0$ modes of our inductively shunted cavity. These modes are formed by EM waves propagating in the x-y plane, and have their electric field oriented along $\hat{\bm{z}}$, for which eq.~(\ref{eq:PlasmaPerm}) applies. Additionally, these modes have EM fields which are independent of $\ell_z$, and so we do not require $\ell_z \rightarrow \infty$. Thus we can apply eqs.~(\ref{eq:PlasmaPerm}) \& (\ref{eq:BelovsFreq}) to the $l=0$ modes of our inductively shunted cavity.
\\
If we define $f_{nm}$ as the $l=0$ mode frequencies of the cavity without the shunt array, and $f'_{nm}$ as the frequencies with the array, then we expect
\begin{eqnarray}
f_{nm}'= \frac{f_{nm}}{\sqrt{\epsilon_{p}(f_{nm}')}}
\label{eq:CavFreq1}
\end{eqnarray}
Inserting eq.~(\ref{eq:PlasmaPerm}), we find the mode frequencies of the inductively shunted cavity to be
\begin{eqnarray}
f_{nm}' = \sqrt{f_{nm}^2+f_{p}^2}
\label{eq:CavFreq2}
\end{eqnarray}
This expression has both a cut-off frequency, the plasma-frequency $f_p$, and also has the desired behaviour of reducing to eq.~(\ref{eq:reccav}) as $r/a \rightarrow 0$. It has previously been used empirically as a fit to FE simulations of cavities containing arrays of thin conducting cylinders \cite{murray2016predicting}. 
\\
We performed HFSS eigenmode simulations to verify eq.~\ref{eq:CavFreq2} over a range of $r/a$, shown in fig.~(\ref{fig:viaLow}), finding good agreement for $r\ll a$. As $r/a$ increases beyond around $0.1$, the model breaks down due to increasing Bragg scattering \cite{pendry1998low}.
%\\
%Eq. \ref{eq:PlasmaPerm} implies that the field distribution of the TM modes is unaffected by the shunts. Apart from directly beside them (where the electric field must be 0), this is borne out in our finite-element simulations, in agreement with previous results \cite{murray2016predicting}.
\\
We can use eq.~(\ref{eq:CavFreq2}) to find the band structure of the cavity in the limit $\ell_x,\ell_y \rightarrow \infty$, by substituting the wavenumbers $ n \pi /\ell_x \rightarrow k_x$, $ m \pi /\ell_y \rightarrow k_y$ into eq.~(\ref{eq:reccav}). On expanding eq.~(\ref{eq:CavFreq2}), we then find the following quadratic mode spectrum near the plasma frequency
\begin{equation}
f = f_p(1+\frac{1}{2}k^2/k_p^2) 
\label{eq:plasmadispersion}
\end{equation}
where $k^2=k_x^2+k_y^2$, $k_p=\sqrt{\epsilon_0\epsilon_r\mu_0}\omega_p$, and $\omega_p=2\pi f_p$.
%%%
\subsection{\label{sec:circuit}Circuit Model $\boldsymbol{(r/a>0.1)}$}
\begin{figure}
  \centering
    \includegraphics[width=0.48\textwidth]{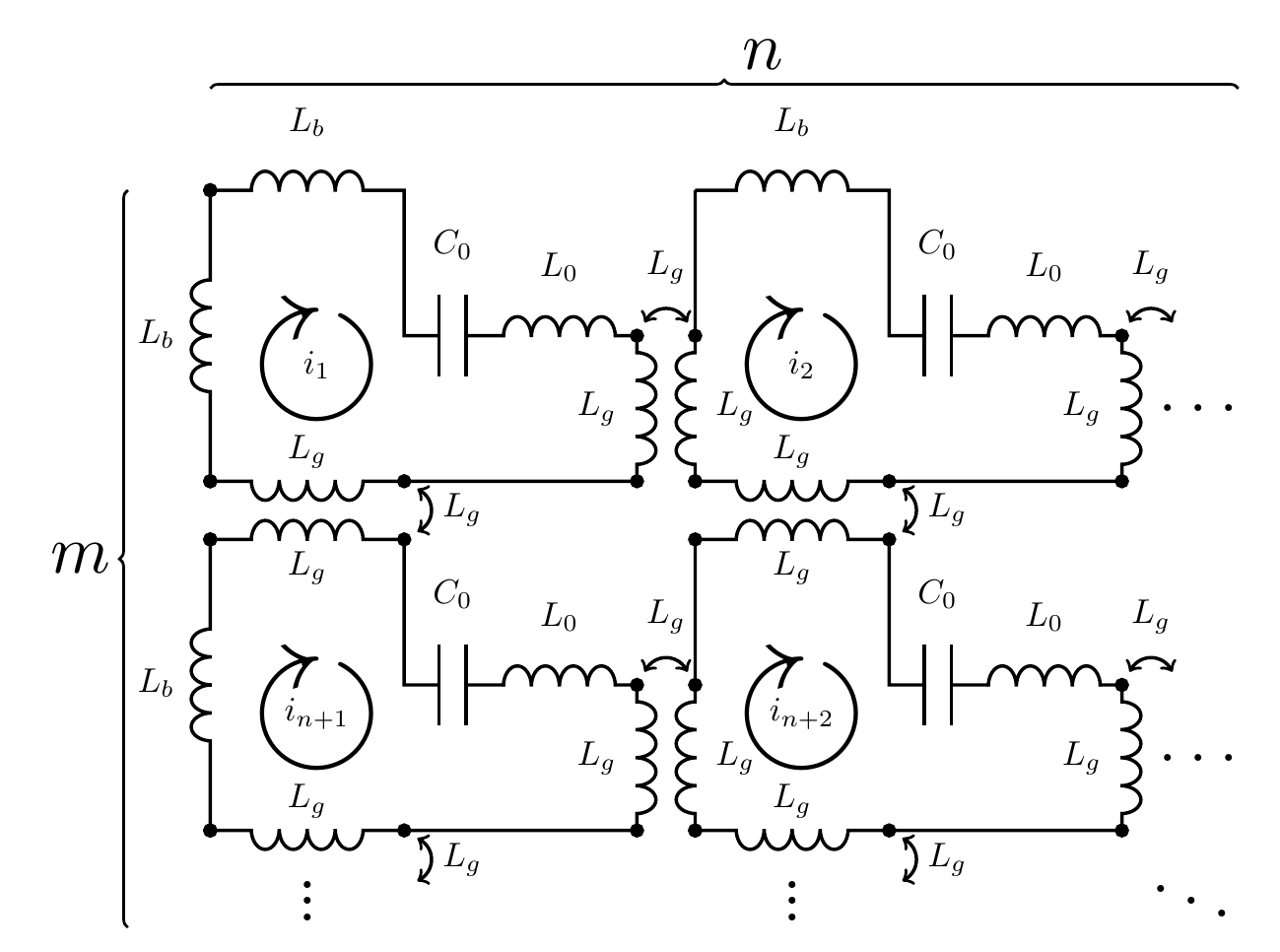}
  \setlength{\belowcaptionskip}{-10pt}
  \vspace{-15pt}
  \caption{Circuit representation for the lowest $n \times m$ modes of a $n \times m$ array of nearest-neighbour magnetically coupled cavities formed between $(n-1) \times (m-1)$ inductive shunts. The isolated fundamental mode of each cavity is represented by $L_0$ and $C_0$. The magnetic coupling between cavities is given by the mutual inductances $L_g$. The circuit includes boundary inductances $L_b$, that can be used to include asymmetry in the outermost cavities. The mutual inductance coupling coefficient $k$ (where $M = k \sqrt{L_1L_2}$) has been set to unity for simplicity, as differences can be absorbed into $L_0$ and $L_b$.}
  \label{fig:2DGrid}
\end{figure}

\begin{figure}
  \centering
    \includegraphics[width=0.48\textwidth]{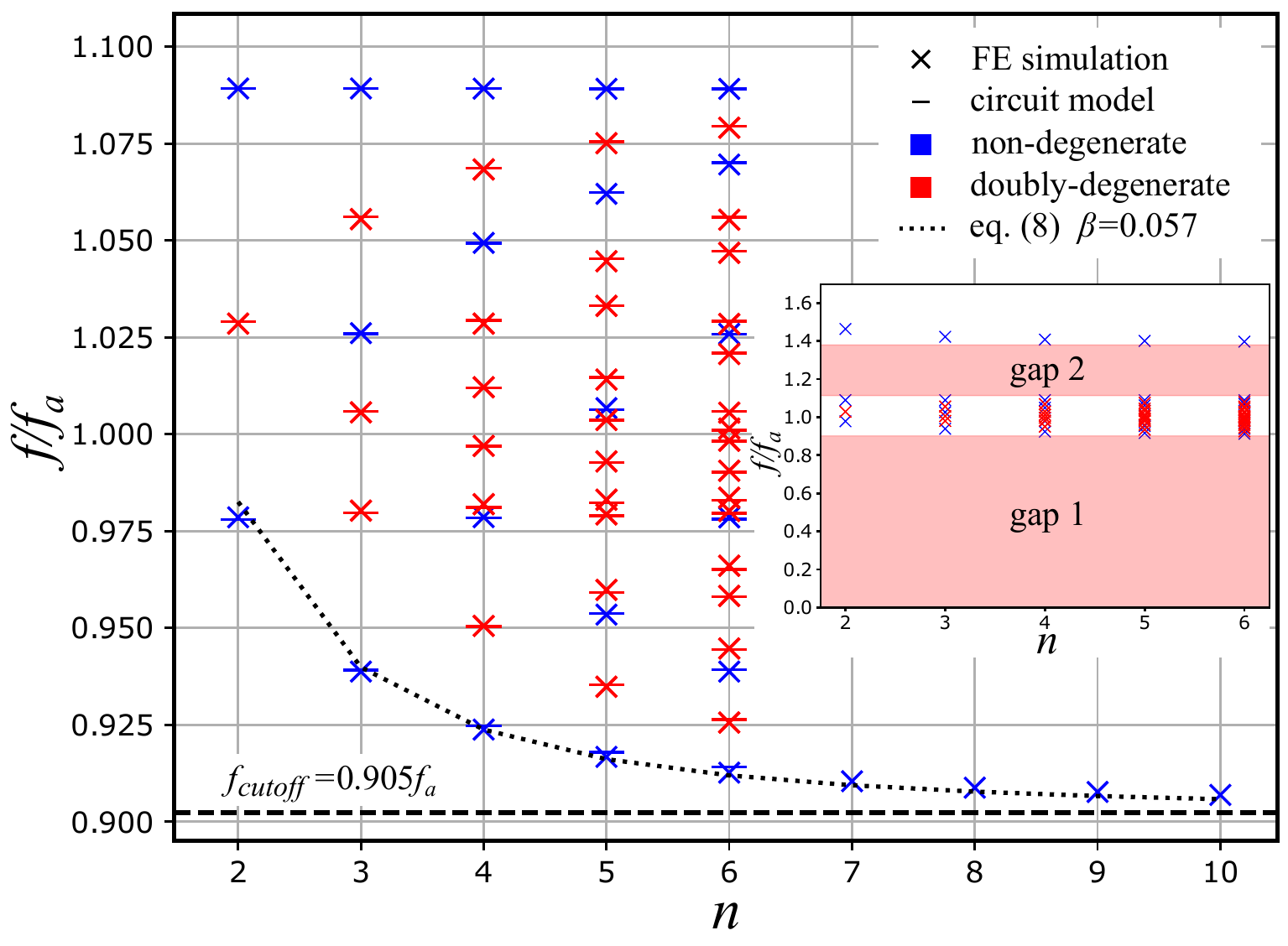}
  \setlength{\belowcaptionskip}{-10pt}
  \vspace{-15pt}
  \caption{Lowest $n^2$ modes for a cavity with size $\ell_x=\ell_y=a \times n$, containing $n-1\times n-1$ inductive shunts, with $r=0.25a$. For n between 2-6, the fit to the next-nearest-neighbour circuit model is shown for all $n^2$ modes, fitted with nearest and next-nearest-neighbour couplings $\beta$ and $\beta_1$ as free parameters. For n between 7-10, only the fundamental frequency from FE simulation is shown. Dotted line shows the fundamental frequency decreasing towards a bound in agreement with equation (\ref{eq:ClosedFormFreq}). Inset shows the lowest $n^2+1$ modes, showing two band-gaps: below mode 1 and between modes $n^2$ and $n^2+1$.}
  \label{fig:VaryingGridSize}
\end{figure}

%%%
In this section, we develop a circuit model for the inductively shunted cavity valid for $r/a>0.1$, where the plasma model has broken down. In this limit, we will model the array of shunts as breaking up the cavity into an array of smaller cavities. We will take the gaps between the shunts into account by allowing neighbouring cavities to magnetically couple to one another.
\\
The tight-binding model has been used to model such coupled-cavity arrays \cite{hartmann2006strongly}, and circuit-models have also been used to model one dimensional coupled-cavity arrays \cite{nagle1967coupled,wangler2008rf}. Here, we extend the circuit-model treatment to two dimensional arrays, and verify that for $r/a>0.1$, it provides an accurate model for the inductively shunted cavity.
\\
Fig.~(\ref{fig:2DGrid}) shows a section of the circuit, from which we construct the impedance matrix $\mathbf{Z_{2D}}$ using mesh analysis. This matrix can be mapped exactly into the simpler impedance matrix $\mathbf{Z_{1D}}$ of a one dimensional coupled-cavity array (See Appendix B for details). For the three specific inductance ratios $L_b/L_g=0, 1, 2$ the circuit has simple closed-form solutions. Taking $L_b=0$ (corresponding to there being no edge effects for cavities at the border of the array), we find the mode frequencies of the inductively-shunted cavity to be
\begin{gather}
f_{ij}= \frac{f_{0}}{\sqrt{1+4 \beta (1+\frac{1}{2}(\cos(\frac{i\pi}{n})+\cos(\frac{j\pi}{m})))}}
\label{eq:ClosedFormFreq}
\\
f_0 = 1/(2\pi \sqrt{L_0 C_0}) \quad \beta=L_g/L_0  \quad (1\leq i \leq n,1\leq j \leq m)
\nonumber
\end{gather}
where $f_{0}$ is the frequency of each uncoupled cavity in the circuit model, and $\beta$ is the inductive coupling parameter between nearest-neighbour cavities. Taking $n,m \rightarrow \infty$, we find a new cutoff frequency for the inductively-shunted cavity,  $f_c=f_{0}/\sqrt{1+8\beta}$.
\\
The field distribution of modes is now significantly altered by the inductive shunt array. We can find the relative field amplitude inside each cavity from the eigenvectors of $\mathbf{Z_{2D}}$. For $L_b=0$, we find
\begin{gather}
 E_{ij}(a,b) = E_0\sin(\frac{i (2a-1)\pi}{2n})\sin(\frac{j (2b-1)\pi}{2m})
 \label{eq:Eigenvectors2D}
 \\
 (1 \leq a,i \leq n) \quad (1 \leq b,j \leq m)
 \nonumber
\end{gather}
where $E_{ij}$ is the relative electric field amplitude in each cavity, $a$ and $b$ index these cavities, and $i$ and $j$ index the modes. The lowest mode ($i,j=1$) is symmetric, and the highest mode ($i=n,j=m$) is anti-symmetric, as we would expect for hybridized modes.
\\
Note that this circuit model tends to the tight-binding model for $\beta \ll 1$. A series expansion of eq.~(\ref{eq:ClosedFormFreq}) in powers of $\beta$ results in
\begin{gather}
f_{ij} \approx f_{0}-2t(2+\cos(k_x a)+\cos(k_y a))
\label{eq:TightBinding}
\\
t=\beta f_{0}/2 \quad k_x = i \pi /\ell_x  \quad k_y = j \pi /\ell_y 
 \nonumber
\end{gather}
which is the tight-binding model dispersion for a square lattice. A series expansion of the cosine terms in eq.~(\ref{eq:ClosedFormFreq}) instead results in the following quadratic mode spectrum near the cut-off frequency
\begin{equation}
f = f_c(1+\frac{1}{2}k^2/k_0^2)
\label{eq:coupledcavquadraticdispersion}
\end{equation}
where $k^2 = k_x^2 + k_y^2$, $k_0^2 = 1/(\beta a^2)$, and $k_x$ and $k_y$ are defined in eq.~(\ref{eq:TightBinding}).
\\
In fig.~(\ref{fig:VaryingGridSize}) we show results of HFSS eigenmode simulations of cavities containing inductive shunt arrays with $r/a=0.25$. We find good agreement to our circuit model, which improves further when we include a next-nearest-neighbour coupling parameter $\beta_1$ (see Appendix B for details).
%%%
\section{\label{sec:applications}INTER-QUBIT COUPLING \& DRIVE-LINE CROSS TALK}
%%%
We now use the results of section \RomanNumeralCaps{2} to predict the form of cavity-mediated cross-talk for superconducting qubits inside enclosures with inductive-shunt arrays. In particular, we consider the transverse exchange coupling $J_{ij}$ between qubits $i$ and $j$, and the drive coupling $\varepsilon_{ij}$ of localized drive-line $i$ to qubit $j$. The corresponding Hamiltonian operators are as follows:
\begin{gather}
\hat{H}^J_{ij} = J_{ij}(\hat{a}_i\hat{a}_j^\dagger+\hat{a}_i^\dagger\hat{a}_j)
\\
\hat{H}^D_{ij} = \varepsilon_{ij} (\hat{a}_j - \hat{a}_j^\dagger) V_i
\end{gather}
Here $\hat{a}^\dagger_i $ and $\hat{a}_i$ are the creation and annihilation operators of qubit mode $i$, $V_i$ is the voltage on drive-line $i$, and `localized drive-line' refers to a drive that interacts with the enclosure from a localized source.
In the absence of any shunts, we expect the mechanism of cavity-mediated cross-talk to be through the standing-wave EM modes of the cavity.
However, in the presence of the inductive shunt array, no standing-waves can form below the cutoff frequency. Instead, the divergence in the density of cavity modes around the cut-off frequency results in a radically different form of cavity-mediated cross-talk. 
\\
The plasma model of periodic inductive shunts provides an intuitive framework for predicting inter-qubit couplings and drive-line cross-talk in this case. In this framework, an excited superconducting qubit or drive-line, oscillating below the plasma frequency, sees the cavity instead as a parallel-plate waveguide below cut-off, and drives an evanescent radial waveguide mode into the cavity. We consider only the dominant $\text{TM}_{00}$ radial mode \cite{marcuvitz1951waveguide}. This will then result in a transverse coupling strength $J_{ij}$ between transmon qubits $i$ and $j$ with the following form (see Appendix C)
\begin{gather}
J_{ij} = 2g^2 \frac{\omega_q}{(v/\delta_0)^2} K_0(d_{ij}/\delta_p)
\label{eq:plasmacouplingstrength}
\\
\delta_p=1/\sqrt{\epsilon_0 \epsilon_r \mu_0 (\omega_p+\omega_q)(\omega_p-\omega_q)}
\label{eq:plasmapendepth}
\end{gather}
where $d_{ij}$ is the qubit separation, $\delta_p$ is the plasma penetration depth, and $g$ is an effective coupling strength between the qubits and the $\text{TM}_{00}$ waveguide mode; $v$ is the speed of light in the waveguide, and $\delta_0$ is the interaction length between qubits and the waveguide; $K_0$ is a modified Bessel function of the second kind, and we have taken qubits $i$ and $j$ to have equal frequency. If the source of the excitation is instead drive-line $i$, this will result in a drive coupling $\varepsilon_{ij}$ to qubit $j$ with the same spatial dependence
\begin{equation}
\varepsilon_{ij} =\varepsilon_0 K_0(d_{ij}/\delta_p)
\label{eq:driveline_coupling}
\end{equation}
where the drive is resonant with qubit $j$. For $d_{ij} \gg \delta_p$, these expressions have simple asymptotic forms, since
\begin{equation}
K_0(d_{ij}/\delta_p) \rightarrow \sqrt{\pi/2} \times  e^{-d_{ij}/\delta_p}/\sqrt{d_{ij}/\delta_p}
\end{equation}
This predicted exponential decay in inter-qubit coupling and drive-line cross-talk is a very useful property where only local qubit connectivity is desired. Note that we can express the plasma penetration depth as a function of only $\omega_q$, $r$ and $a$
\begin{gather}
\delta_p=a\sqrt{(\ln (\frac{a}{r})-\Pi)/2\pi}\sqrt{1/(1-(\omega_q/\omega_p)^2)}
\label{eq:plasmadepthexplicit}
\end{gather}
where $\omega_p$ is itself a function of $r$ and $a$ given by eq.~(\ref{eq:BelovsFreq}), and where $\Pi$ is also defined in eq.~(\ref{eq:BelovsFreq}).
\\
We can instead derive the cavity-mediated coupling between qubits by considering the interaction of qubits with the new distribution of cavity modes in the inductively shunted cavity. From eqs.~(\ref{eq:plasmadispersion}) \& (\ref{eq:coupledcavquadraticdispersion}), we see that the inductively shunted cavity is characterised by a 2D quadratic mode spectrum above the cut-off frequency. A qubit with a frequency below the cut-off will interact with these modes to form a bound state \cite{shi2016bound}, with a spatially exponentially decaying envelope. These bound states will then mediate a coupling between distant qubits \cite{douglas2015quantum}. Remarkably, for equal frequency qubits below the cut-off of a 2D quadratic mode spectrum, the predicted bound-state mediated transverse coupling \cite{douglas2015quantum,gonzalez2015subwavelength} has exactly the same spatial dependence as eq.~(\ref{eq:plasmacouplingstrength}), with the plasma penetration depth replaced by the bound state length
\begin{equation}
\delta_b = \sqrt{\alpha \omega_b/(\omega_b-\omega_q)}
\label{eq:boundstates_length}
\end{equation}
where $\alpha$ characterises the curvature of the band-edge, given by $\omega = \omega_b(1+\alpha(k-k_0)^2)$.
\begin{table*}
\caption{\label{tab:mode_freqs} fundamental frequency of the enclosure in figure~\ref{fig:monolithicdevicediagram} (\SI{}{\GHz}), changing the inductive shunt radius (\SI{}{\mm}).}
\begin{ruledtabular}
\begin{tabular}{lccccccccc}
\textrm{}&
\textrm{$0$\footnote{no inductive shunts}}&
\textrm{$0.05$}&
\multicolumn{1}{c}{\textrm{$0.1$}}&
\textrm{$0.15$}&
\textrm{$0.2$}&
\textrm{$0.25$}&
\textrm{$0.3$}&
\textrm{$0.35$}&
\textrm{$0.4$} \\
\colrule
FE\footnote{HFSS Finite Element eigenmode simulation} & 1.49 & 11.89 & 13.95 & 15.78 & 17.58 & 19.42 & 21.37 & 23.44 & 25.14 \\
Eq.~(\ref{eq:CavFreq2}) & 1.46 & 11.34 & 13.43 & 15.39 & 17.47 & 19.82 & 22.68 & 26.40 & 31.74 \\
\end{tabular}
\end{ruledtabular}
\end{table*}
We can find $\alpha$ from eq.~(\ref{eq:plasmadispersion}) and compare the predicted bound state length with the plasma penetration depth in eq.~(\ref{eq:plasmapendepth}). We find
\begin{equation}
\delta_b = 1/\sqrt{\epsilon_0 \epsilon_r \mu_0 2\omega_p(\omega_p-\omega_q)}
\end{equation}
This result will hold for qubits close in frequency to the band-edge, where the quadratic dispersion approximation is valid; in which case $\omega_q \approx \omega_p$ and $\delta_b \approx \delta_p$. Thus the spatial dependence of qubit coupling for these two seemingly disparate models agrees well. Finding $\alpha$ instead from eq. (\ref{eq:coupledcavquadraticdispersion}), we find
\begin{equation}
\delta_b = a\sqrt{\beta/2}\sqrt{1/(1-\omega_q/\omega_b)}
\label{eq:circuitmodel_boundstatelength}
\end{equation}
Note that this last result applies not just to inductively shunted cavities, but more generally to 2D coupled cavity arrays.
%%%
\section{\label{sec:simulation}FE SIMULATION OF MONOLITHIC SUPERCONDUCTING QUBIT DEVICE}
\begin{figure*}
  \centering
    \includegraphics[width=0.9\textwidth]{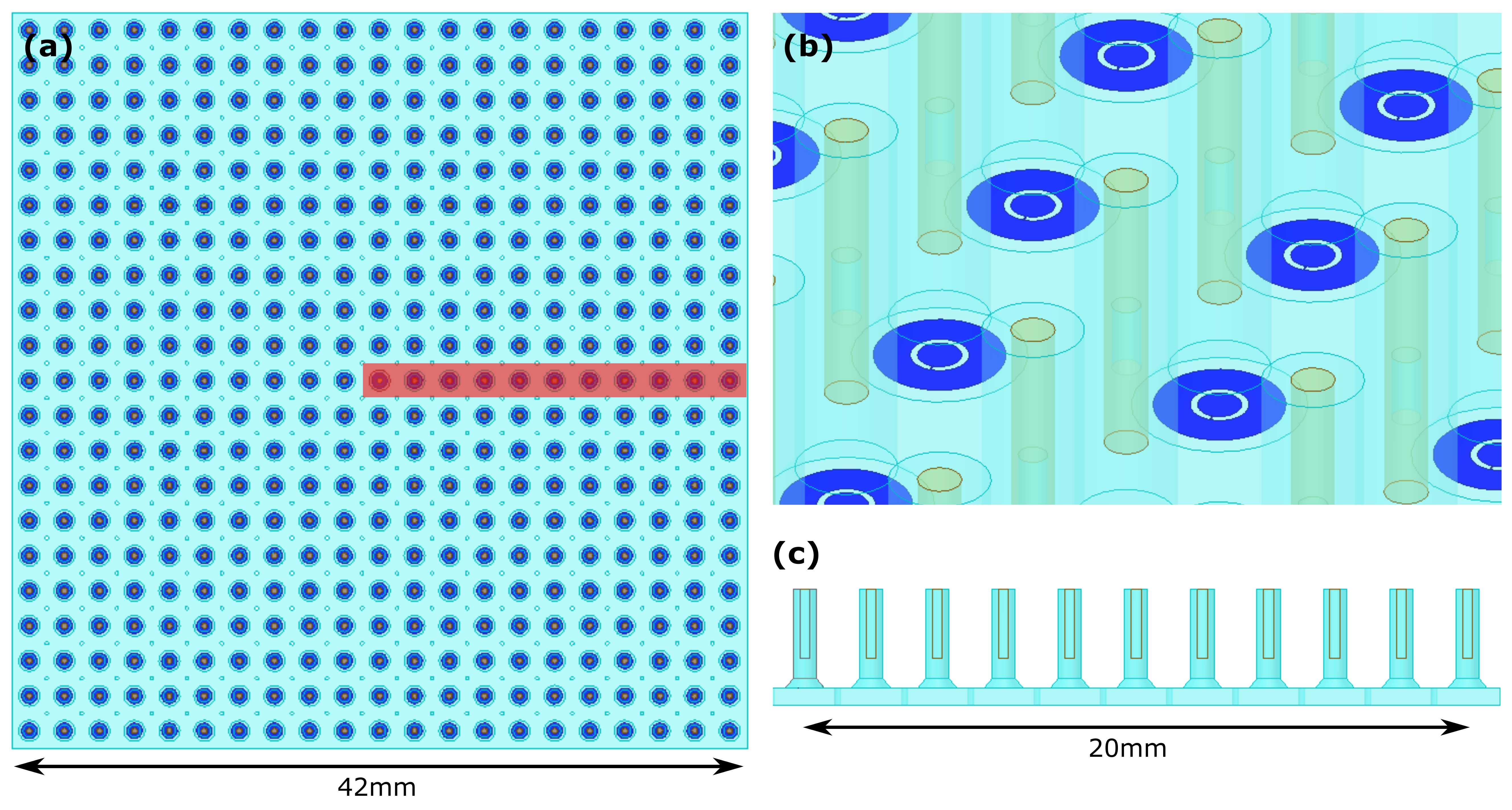}
  \setlength{\belowcaptionskip}{-10pt}
  \vspace{-15pt}
  \caption{\textbf{(a)} Top-down view of the model. The qubits for which $J$ and drive-line cross-talk were simulated are shaded red. \textbf{(b)} Trimetric-view of the model. The qubit islands are shaded blue, and the coaxial drive-lines are shaded yellow. These qubits have $C_J \approx 100$fF and $L_J \approx 10$nH at $f_{01}=5$GHz, leading to $E_J/E_C \approx 80$. \textbf{(c)} Cross-section of the region shaded in \textbf{(a)}. The model is enclosed by a perfectly conducting boundary to represent an ideal superconducting enclosure.}
  \label{fig:monolithicdevicediagram}
\end{figure*}

\begin{figure*}
  \centering
    \includegraphics[width=0.9\textwidth]{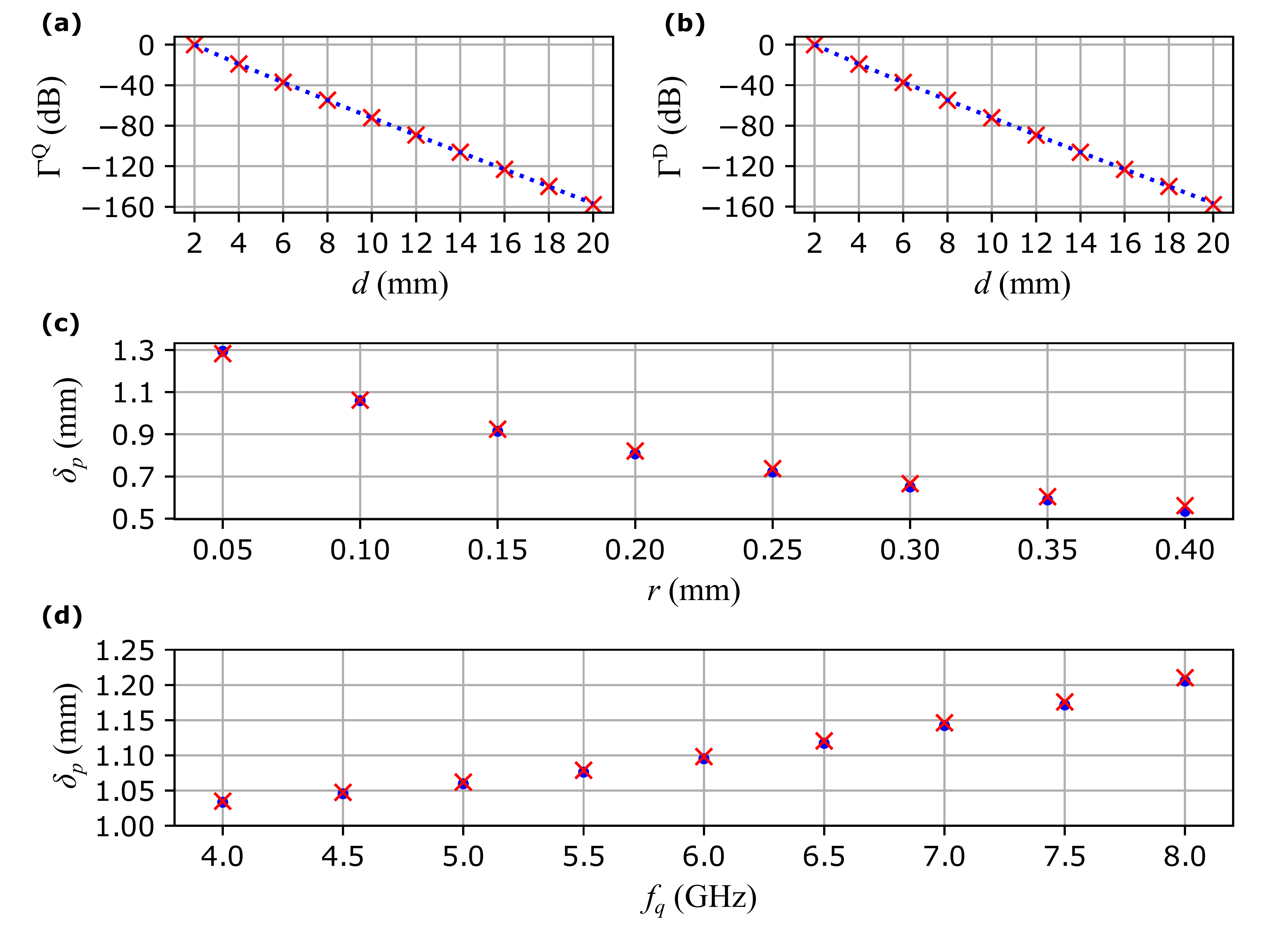}
  \setlength{\belowcaptionskip}{-10pt}
  \vspace{-15pt}
  \caption{\textbf{(a)} and \textbf{(b)} Inter-qubit and drive-line to qubit cross-talk for $r=0.1$mm, $f_q = 5$GHz. Red crosses are FE simulation values, and blue dotted lines are the prediction of eqs.~(\ref{eq:plasmacouplingstrength}), (\ref{eq:plasmapendepth}), (\ref{eq:driveline_coupling}). \textbf{(c)} and \textbf{(d)} Plasma penetration depth against shunt radius and qubit frequency. Red crosses are calculated using eq.~(\ref{eq:plasmapendepth}) and blue dots are found by fitting eq.~(\ref{eq:plasmacouplingstrength}) to the FE simulation $\Gamma^Q$ values, with $\delta_p$ as the sole fit parameter. In \textbf{(c)} $f_q=5$GHz, and in \textbf{(d)} $r=0.1$mm.}
  \label{fig:inter-qubitcouplingfits}
\end{figure*}

%%%
We now move on to perform FE simulations of a realistic superconducting circuit device, and compare the results with the predictions of the previous sections. We performed HFSS simulations on the superconducting qubit device model shown in fig.~(\ref{fig:monolithicdevicediagram}). This model is based on an architecture in which universal quantum control and readout have been demonstrated \cite{rahamim2017double,patterson2019calibration}. For the purpose of probing cavity-mediated cross-talk qubit couplers and readout circuitry are not included in the model. The model consists of a $21\times21$ array of coaxial transmon qubit islands on a silicon ($\epsilon_r=11.9$) substrate measuring $\ell_x=\SI{42}{\mm}$, $\ell_y=\SI{42}{\mm}$, $\ell_z=\SI{0.5}{\mm}$, enclosed by a perfectly conducting cavity, which is inductively shunted by a $20\times20$ array of perfectly conducting cylinders. The qubit islands and cylinders are spaced by $a=\SI{2}{\mm}$, and each pair of qubit islands is capacitively coupled to an off-chip coaxial drive-line.
\\
The fundamental mode frequency of the cavity over a range of cylinder radii is summarised in Table~\ref{tab:mode_freqs}. In the absence of any cylinders, the fundamental frequency of the enclosure is well below the typical range of transmon frequencies, which would result in a high density of enclosure modes around qubit frequencies. However, in the presence of the considered shunting arrays, the fundamental mode frequency is in all cases greater than \SI{11}{\GHz}. For $r/a>0.1$, eq.~(\ref{eq:CavFreq2}) rapidly diverges from the simulation result as it breaks down due to Bragg scattering. For $r/a<0.1$, we attribute the difference to the vacuum regions in the simulation model introduced by the drive-lines, which are not included in eq.~(\ref{eq:CavFreq2}).
\\
We defined a reference qubit (drive) $i=0$ at the centre of the array, and simulated the transverse coupling (drive-line cross-talk) to qubit $j={1,2,...,10}$ using a simple impedance formula \cite{solgun2019simple}. While simulating these properties between qubit (drive) $i$ and qubit $j$, the Josephson junctions in all other qubits are replaced by open circuits. This ensures we only simulate cross-talk effects coming from the enclosure, while also simplifying simulation complexity.
\\
We re-express our results using
\begin{gather}
\Gamma_{0,j}^Q =J_{0,j}/J_{0,1}
\\
\Gamma_{0,j}^D = \varepsilon_{0,j}/\varepsilon_{0,1}
\end{gather}
These expressions eliminate the unknown prefactors in eqs.~(\ref{eq:plasmacouplingstrength}) \&  (\ref{eq:driveline_coupling}) considering qubits which couple equally to the evanescent waveguide mode. In this case $\Gamma^Q=\Gamma^D$, since both are then only a measure of the spatial decay of the mediating waveguide mode. Ignoring the weak square-root dependence on qubit-separation, these expressions approximate to
\begin{gather}
\Gamma_{0,j}^Q = \Gamma_{0,j}^D \approx e^{-d_{1,j}/\delta_p}
\end{gather}
We find good agreement between simulation and the plasma-model of inter-qubit coupling and drive-line cross-talk for a range of qubit frequencies and shunt radii (see fig.~\ref{fig:inter-qubitcouplingfits}). When calculating the plasma penetration depth (eq.~\ref{eq:plasmapendepth}), we used the fundamental cavity frequency found from eigenmode simulation (Table~\ref{tab:mode_freqs}) as the plasma frequency, rather than that from eq.~(\ref{eq:BelovsFreq}). With this choice, we find that the plasma model of inter-qubit coupling still agrees well with simulation for $r/a>0.1$.
\\
The plasma penetration depth is of the order of the shunt separation, and decreases significantly with increasing shunt radius. For qubits well below the plasma frequency, it has only a weak dependence on the qubit frequency.
\\
Note that although we have used coaxial qubits in our simulation, we can expect the qubit geometry only to affect the coupling $g$ to the dominant $\text{TM}_{00}$ waveguide mode, and therefore expect good agreement with the model for other qubit geometries.
%%%
\section{CONCLUSIONS}
%%%
We have first developed a plasma and circuit model to accurately predict the mode frequencies of enclosures inductively shunted by periodic perfectly-conducting cylinder arrays. We have then used these models to predict the exponential decay of cavity-mediated inter-qubit coupling and drive-line cross-talk for superconducting circuits inside such enclosures. The plasma model in particular predicts the fundamental enclosure frequency and the rate of cross-talk decay to be simple functions of the shunt radius and spacing (eqs. \ref{eq:CavFreq2} \& \ref{eq:plasmadepthexplicit}), providing a tool for the design of the shunt array. The predictions agree well with a FE simulation of a realistic device. These results indicate that monolithic superconducting circuit architectures that employ inductive shunt arrays can scale arbitrarily in size, with enclosure-mediated cross-talk that is small and local in nature, making this a promising approach for quantum computation with superconducting circuits.
%%%
\begin{acknowledgments}
P.S. acknowledges support from NQIT (Networked Quantum Information Technologies). T.T. acknowledges support from the Masason foundation and the Nakajima Foundation. B.V. acknowledges support from an EU Marie Skodowska-Curie fellowship. P.L. acknowledges support from the EPSRC
[EP/M013243/1] and Oxford Quantum Circuits Limited. We thank C. Murray for insightful discussions, and S. Sosnina for technical contributions.
\end{acknowledgments}
%\appendix
\section*{APPENDIX A: EFFECT OF MULTIPLE DIELECTRIC LAYERS}
\begin{figure}[H]
  \centering
    \includegraphics[width=0.48\textwidth]{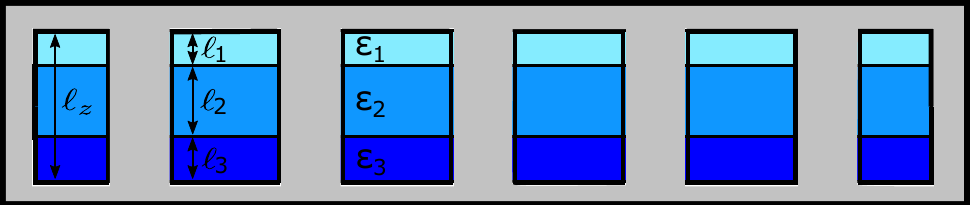}
  \caption{Cross-section of the inductively shunted cavity, now containing a stack of different dielectric materials.}
  \label{fig:stackeddielectrics}
\end{figure}

Here we consider replacing the single dielectric in the cavity with multiple layers of dielectric as in fig.~(\ref{fig:stackeddielectrics}). A relevant case for superconducting quantum circuits is that with three layers: vacuum, substrate, vacuum.
\\
Since the magnetic properties of the cavity are unaltered, the introduction of multiple dielectric layers will only affect the capacitance of the $l=0$ modes, which take the form of a parallel plate capacitance between the top and bottom of the cavity
\begin{eqnarray}
C  = k \times \frac{\epsilon_r}{\ell_z}
\end{eqnarray}
where $k$ is a constant. The capacitance in the presence of multiple dielectric layers $C'$ is the series sum of the parallel plate capacitances across each layer
\begin{gather}
C' = k \times 1/\Sigma_{i=1}^n(\frac{\ell_i}{\epsilon_i})=k\times \frac{\epsilon'_r}{\ell_z}
\\
\epsilon'_r = \ell_z/\Sigma_{i=1}^n(\frac{\ell_i}{\epsilon_i})
\label{eq:effectivepermitivitty}
\end{gather}
Thus the only effect of a dielectric stack is to replace the relative permitivitty $\epsilon_r$ wherever it appears, with the effective relative permitivitty $\epsilon'_r$ in eq.~(\ref{eq:effectivepermitivitty}).
\section*{APPENDIX B: MESH ANALYSIS \& MAPPING THE ARRAY CIRCUIT TO CHAIN CIRCUIT}
\begin{figure}
  \centering
    \includegraphics[width=0.5\textwidth]{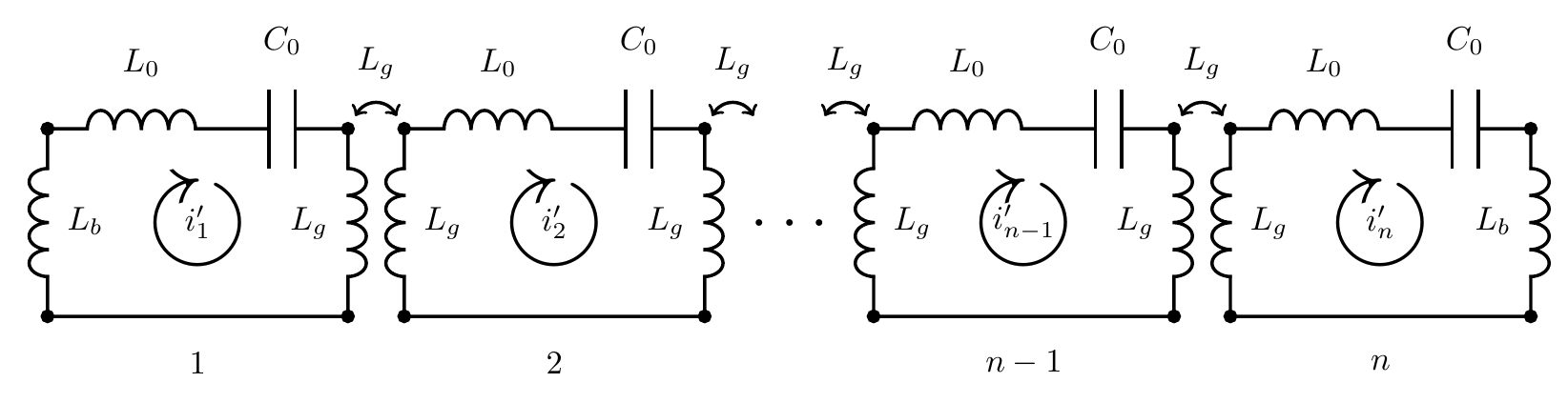}
  \setlength{\belowcaptionskip}{-10pt}
    \vspace{-15pt}
  \caption{Circuit representation for the lowest $n$ modes of a chain of $n$ magnetically coupled cavities.}
  \label{fig:1DChain}
\end{figure}

Using mesh analysis \cite{hayt1978engineering}, the circuit in fig.~(\ref{fig:2DGrid}) of the main text can be represented by
\begin{eqnarray}
\mathbf{Z_{2D}} \begin{bmatrix}  
  i_{1}  \\
   \vdots           \\
   i_{i}    \\
   \vdots    \\
   i_{nm} \end{bmatrix} =
\begin{bmatrix}
  V_{1}  \\
                \vdots           \\
    V_{i}               \\
         \vdots          \\
V_{nm}
\end{bmatrix}
\end{eqnarray}
Where $\mathbf{Z_{2D}}$ is an $nm\times nm$ square matrix, $i_i$ is the current through mesh $i$ and $V_i$ is the voltage applied to mesh $i$. At the mode frequencies, currents can oscillate in the absence of any excitations ($V_i=0$), therefore modes exist at frequencies where an eigenvalue of $\mathbf{Z_{2D}}$ is $0$.
\\
$\mathbf{Z_{2D}}$ can be expressed as a diagonal block matrix. For a nearest-neighbour coupling model this block matrix is tridiagonal, while for a next-nearest-neighbour coupling model it is pentadiagonal, and so on. We will consider the cases of nearest-neighour and next-nearest-neighbour couplings. Expressed in block form, $\mathbf{Z_{2D}}$ is a $m \times m$ matrix of $n \times n$ matricies. For nearest-neighbour coupling, it takes the form
\begin{eqnarray*}
\mathbf{Z_{2D_{m \times m}}} = \begin{bmatrix}
  \mathbf{Z\alpha} & \mathbf{Z_G} &                      &            &  \cdots  &      &         \mathbf{0} \\
  \mathbf{Z_G} & \mathbf{Z\beta} & \mathbf{Z_G} &                &   &                                   &              \\
  &\mathbf{Z_G} & \mathbf{Z\beta} & \mathbf{Z_G} &                &                         &    \\
 &  &\ddots&\ddots&\ddots& &              \\
  & &  & \mathbf{Z_G} & \mathbf{Z\beta} &\mathbf{Z_G} &\\ 
  & & &  &\mathbf{Z_G} &\mathbf{Z\beta} &\mathbf{Z_G}\\      
 \mathbf{0}& & \cdots & &  & \mathbf{Z_G} & \mathbf{Z\alpha}\\      
\end{bmatrix}
\end{eqnarray*}
where 
\begin{eqnarray*}
\mathbf{Z \alpha_{n\times n}} = \begin{bmatrix}
  Z_{\alpha_1}& -Z_g &                      &            &  \cdots  &      &         0 \\
  -Z_g & Z_{\alpha_2} & Z_b &                &   &                                   &              \\
  & -Z_g & Z_{\alpha_2} & -Z_g &                &                         &    \\
 &  &\ddots&\ddots&\ddots& &              \\
  & &  & -Z_g & Z_{\alpha_2} & -Z_g &\\ 
  & & &  & -Z_g & Z_{\alpha_2} & -Z_g\\      
 0& & \cdots & &  & -Z_g & Z_{\alpha_1}\\      
\end{bmatrix}
\end{eqnarray*}
\begin{align*}
&Z_{\alpha_1}=Z_0+2Z_g+2Z_b\\
&Z_{\alpha_2}=Z_0+3Z_g+Z_b
\end{align*}
\begin{eqnarray*}
\mathbf{Z \beta_{n \times n}} = \begin{bmatrix}
  Z_{\beta_1}& -Z_g &                      &            &  \cdots  &      &         0 \\
  -Z_g & Z_{\beta_2} & Z_b &                &   &                                   &              \\
  & -Z_g & Z_{\beta_2} & -Z_g &                &                         &    \\
 &  &\ddots&\ddots&\ddots& &              \\
  & &  & -Z_g & Z_{\beta_2} & -Z_g &\\ 
  & & &  & -Z_g & Z_{\beta_2} & -Z_g\\      
 0& & \cdots & &  & -Z_g & Z_{\beta_1}\\      
\end{bmatrix}
\end{eqnarray*}
\begin{align*}
&Z_{\beta_1}=Z_0+3Z_g+Z_b\\
&Z_{\beta_2}=Z_0+4Z_g
\end{align*}
\begin{eqnarray*}
\mathbf{Z_{G_{n \times n}}} = -Z_g\mathbf{I_{n \times n}}
\end{eqnarray*}
and
\begin{eqnarray*}
Z_0 = i\omega L_0-\frac{i}{\omega C_0} \quad \quad Z_g = i \omega L_g \quad \quad Z_b = i \omega L_b
\end{eqnarray*}
It can be verified by expansion that $\mathbf{Z_{2D}}$ can be written
\begin{eqnarray}
\mathbf{Z_{2D_{nm \times nm}}} = \mathbf{Z_{1D_{n\times n}}} \oplus \mathbf{Z_{1D_{m \times m}}} - \scaleto{Z_{0}}{8pt}\mathbf{I_{nm \times nm}}
\label{eq:2dto1dmap}
\end{eqnarray}
where $\oplus$ is the Kronecker sum and 
\begin{eqnarray*}
\mathbf{Z_{1D_{n\times n}}} =\begin{bmatrix}
 Z_1 &    -Z_g  &          &                        \cdots &                           0 \\
 -Z_g & Z_2 & -Z_g              &                  &                  &                  \\
                 &                \ddots &  \ddots &  \ddots &                  &                  \\     
                                            &             & -Z_g&  Z_2& -Z_g \\
               0 &                      \cdots               &               &   -Z_g & Z_1
\end{bmatrix}
\end{eqnarray*}
\begin{align*}
&Z_1=Z_0+Z_g+Z_b\\
&Z_2=Z_0+2Z_g
\end{align*}
$\mathbf{Z_{1D_{n\times n}}}$ is exactly the impedance matrix of the circuit in fig.~(\ref{fig:1DChain}). This mapping makes solving the eigenvalues of $\mathbf{Z_{2D}}$ much easier, since
\begin{eqnarray}
\lambda(\mathbf{Z_{2D}})_{ij} = \lambda(\mathbf{Z_{1D}})_{i} +  \lambda(\mathbf{Z_{1D}})_{j} -Z_0   
\label{eq:eigenvalsmap}
\end{eqnarray}
Therefore, finding the mode frequencies of the 2D circuit is reduced to the problem of solving the eigenvalues of the 1D circuit, inserting these into eq.~(\ref{eq:eigenvalsmap}), and solving for $\lambda(\mathbf{Z_{2D}})_{ij} = 0$.
\\
$\mathbf{Z_{1D}}$ is a tridiagonal matrix that is also almost a toeplitz matrix except for the elements $\mathbf{Z_{1D}}_{11}$ and $\mathbf{Z_{1D}}_{nn}$. Simple closed-form solutions for this type of matrix are known to exist for particular values of these boundary elements \cite{losonczi1992eigenvalues}, in our case when $L_b = 0,L_g,2L_g$. Using these solutions and substituting into eq.~(\ref{eq:eigenvalsmap}) leads to the following mode frequencies for the circuit in fig.~(\ref{fig:2DGrid}) of the main text
\begin{eqnarray}
f_{ij}= \frac{f_{0}}{\sqrt{1+4 \beta (1+\frac{1}{2}\gamma_{ij})}} \quad \quad
\\
\nonumber
\gamma_{ij} = \cos(\frac{i\pi}{n})+\cos(\frac{j\pi}{m}) \quad (L_b=0)
\\
\nonumber
\gamma_{ij} = \cos(\frac{j\pi}{n+1})+\cos(\frac{j\pi}{m+1}) \quad (L_b=L_g)
\\
\nonumber
\gamma_{ij}=\cos(\frac{(i-1)\pi}{n})+\cos(\frac{(j-1)\pi}{m}) \quad (L_b=2L_g)
%f_{ij_{L_b=0}}= \frac{f_{0}}{\sqrt{1+4 \beta %(1+\frac{1}{2}(\cos(\frac{i\pi}{n})+\cos(\frac{j\p%i}{m})))}}
%\\
%f_{ij_{L_b=L_g}}= \frac{f_{0}}{\sqrt{1+4 \beta %(1+\frac{1}{2}(\cos(\frac{i\pi}{n+1})+\cos(\frac{j%\pi}{m+1})))}}
%\\
%f_{ij_{L_b=2L_g}}= \frac{f_{0}}{\sqrt{1+4 \beta %(1+\frac{1}{2}(\cos(\frac{(i-1)\pi}{n})+\cos(\frac%{(j-1)\pi}{m})))}}
\end{eqnarray}
Increasing the boundary inductance $L_b$ lowers the frequency of all modes in these solutions.
\\
We find it interesting that the 2D circuit can be mapped directly into the far simpler 1D circuit. We note that this is only true for the special case of identical inductances and capacitances across all unit cells, and that in general, this mapping is not possible.
\begin{figure}
  \centering
    \includegraphics[width=0.48\textwidth]{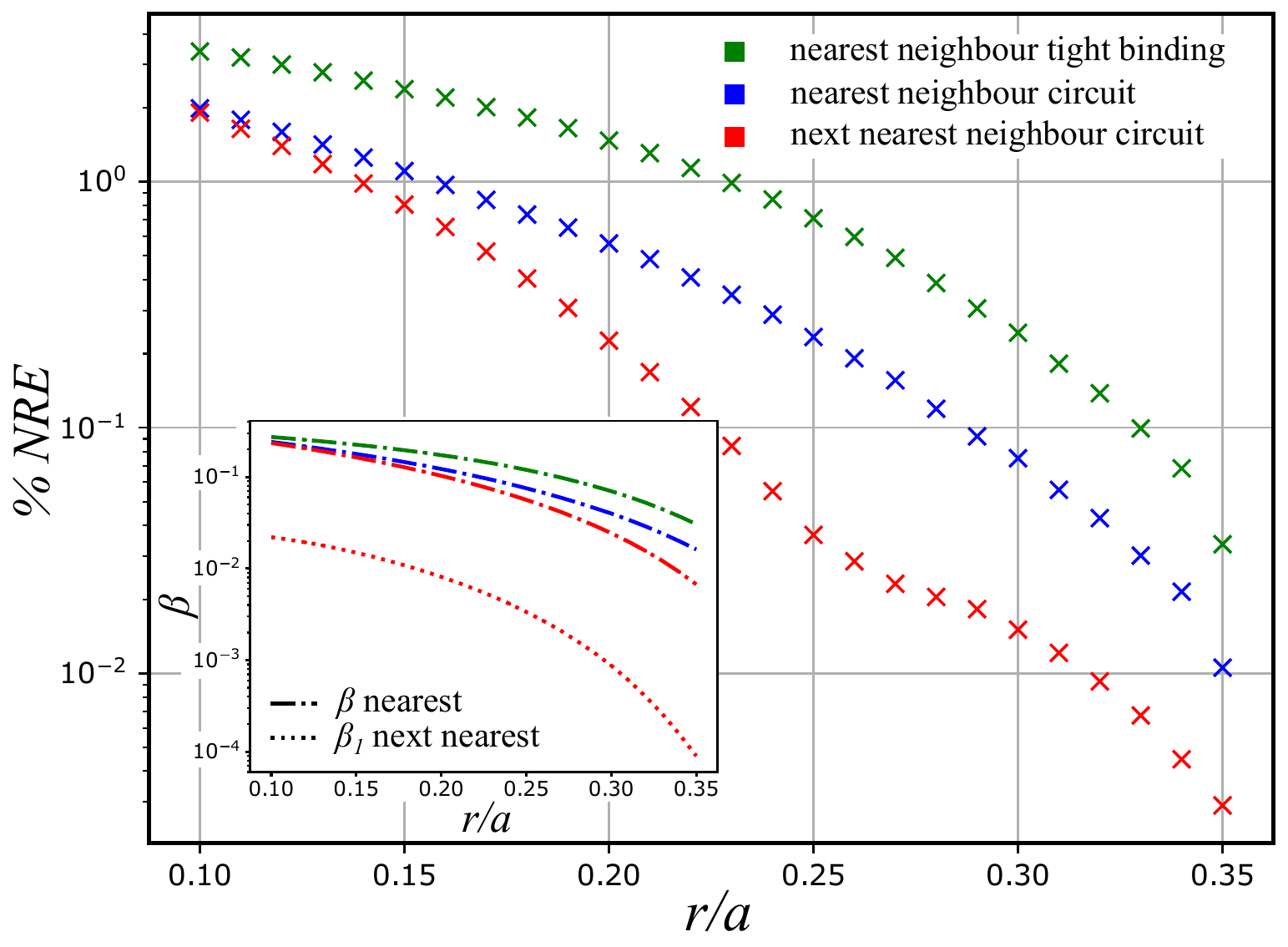}
  \setlength{\belowcaptionskip}{-10pt}
  \vspace{-20pt}
  \caption{Normalized relative error (NRE) $\Sigma_{i=1}^{16}|f_{i_{\text{FE}}}-f_{i_{\text{model}}}| /16f_{i_\text{FE}}$, between a FE simulation and three different fit models, for the lowest 16 modes of a cavity containing a $3\times 3$ inductive shunt array, changing shunt radius $r$. Inset shows the fitted coupling parameters decreasing smoothly with increasing inductive shunt radius.}
  \label{fig:Residuals_Circ}
\end{figure}
\\
%%%%%% pentadiagonal %%%%%%
We now consider the next-nearest-neighbour coupling case, to demonstrate the effect of including further couplings. In this case, it can again be verified by expansion that $\mathbf{Z_{2D}}$ can be written in the same form as eq.~(\ref{eq:2dto1dmap}), where now
\begin{eqnarray*}
\mathbf{Z_{1D_{n \times n}}} =\begin{bmatrix}
 Z_1 &    -Z_g  &-Z_{g_2}           &                &           \cdots &                  &                0 \\
 -Z_g & Z_2 & -Z_g &  -Z_{g_2}              &                  &                  &                  \\
 -Z_{g_2}&-Z_g & Z_3 & -Z_g &  -Z_{g_2}              &                  &                    \\
                 &                & \ddots &  \ddots &  \ddots &                  &                  \\             \\
                          &                &         -Z_{g_2}       & -Z_g&  Z_3& -Z_g&-Z_{g_2} \\
                 &                &                &         -Z_{g_2}       & -Z_g&  Z_2& -Z_g \\
               0 &                &         \cdots &                &              -Z_{g_2}    &   -Z_g & Z_1
\end{bmatrix}
\end{eqnarray*}
\begin{align*}
&Z_1=Z_0+Z_g+Z_{g_2}+Z_b\\
&Z_2=Z_0+2Z_g+Z_{g_2}\\
&Z_3=Z_0+2Z_g+2Z_{g_2}.
\end{align*}
%\begin{figure}
%  \centering
%    \includegraphics[width=0.45\textwidth]{figures/Appendix_closedformfreqs.pdf}
%      \setlength{\belowcaptionskip}{-10pt}
%    \vspace{-15pt}
%  \caption{Mode frequencies for a $10 \times 10$ array, with $\beta=0.05$. Increasing the boundary inductance $L_b$ lowers the frequency of all modes.}
%  \label{fig:ClosedFormFreqs}
%\end{figure}
This is again the impedance matrix of the circuit in fig.~(\ref{fig:1DChain}), now with next nearest-neighbour couplings switched on. Simple closed form solutions to $\mathbf{Z_{1D_{n \times n}}}$ no longer exist in this case, however we have still greatly reduced the problem from one of finding the eigenvalues of a $\mathbf{mn} \times \mathbf{mn}$ matrix to one of finding the eigenvalues of an $\mathbf{m \times m}$ matrix and an $\mathbf{n \times n}$ matrix. 
\\
The agreement of the nearest and next nearest neighbour circuit models, as well as the tight binding model in eq.~(\ref{eq:TightBinding}), to a FE simulation are shown in fig.~(\ref{fig:Residuals_Circ}). The agreement for all models increases as $r/a$ increases. The circuit model offers a better fit than the tight binding model for a single free parameter.
\section*{APPENDIX C: TRANSVERSE COUPLING IN PLASMA MODEL}
\begin{figure}
  \centering
    \includegraphics[width=0.45\textwidth]{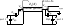}
      \setlength{\belowcaptionskip}{-15pt}
    \vspace{-15pt}
  \caption{Circuit model for two transmon qubits (with junctions replaced by ports) separated by distance $d_{12}$ and coupled by the evanescent $\text{TM}_{00}$ mode of a radial waveguide.}
  \label{fig:PlasmaCircuitModel}
\end{figure}

To find the transverse coupling between the two qubits, we will use the impedance formula presented in Ref.  \cite{solgun2019simple}
\begin{equation}
    J_{ij} = -1/4\sqrt{\frac{\omega_i\omega_j}{L_iL_j}}\text{Im}[\frac{Z_{ij}(\omega_i)}{\omega_i}+\frac{Z_{ij}(\omega_j)}{\omega_j}]
    \label{eq:JImpedanceFormula}
\end{equation}
This expression, valid for weakly anharmonic transmon qubits, reduces the problem of finding $J_{ij}$ to that of finding the trans-impedance $Z_{ij}(\omega)$ between ports $i$ and $j$, which replace the Josephson-junctions of qubits $i$ and $j$. $\omega_{i}$ is the frequency of qubit $i$, and $L_{i}$ is closely related to the bare junction inductance of qubit $i$ \cite{solgun2019simple}.
\\
To find the trans-impedance  between qubits inside the inductively shunted cavity, we use the circuit model in fig.~(\ref{fig:PlasmaCircuitModel}). Two transmon qubits, with their Josephson-Junctions replaced by ports, are each capacitively coupled to a parallel-plate waveguide, with $\delta_0$ representing the interaction length between the qubits and the waveguide. A current is driven through the port of qubit 1, which will drive a radial waveguide mode, centred around qubit 1, into the waveguide. This mode will propagate out to qubit 2, a distance $d_{12}$ away, where it will induce a voltage across port 2. The expression for the incident voltage wave along the waveguide for the dominant $\text{TM}_{00}$ mode is \cite{marcuvitz1951waveguide}
\begin{equation}
   v_i(d) = a \times v_{in}H^{(2)}_0(kd)
    \label{eq:incident}
\end{equation}
%\begin{gather}
%    v_i(d) \approx v_{in}e^{ikd}/\sqrt{kd}
 %   \label{eq:incident}
 %   \\
 %   v_r(d) \approx \Gamma_Rv_i(d_{12})e^{ik(d_{12}-d)}/\sqrt{ k(d_{12}-d)}
 %   \label{eq:reflected}
 %   \end{gather}
$k$ is the wavenumber of the line, $a$ is a normalisation factor $a=1/H^{(2)}_0(k\delta_0)$ and $H_0^{(2)}$ is the Hankel function of the second kind. The current wave has a similar form.
\\
The wavenumber is given by
\begin{equation}
    k = \sqrt{\mu \epsilon_0 \epsilon_r \epsilon_p}\omega
\end{equation}
where $\epsilon_p$ is given by eq.~(3) in the main text, and we have used that the electric field is parallel to $z$ for the $\text{TM}_{00}$ mode. $k$ is therefore imaginary at frequencies below the plasma frequency, and the voltage and current waves through the waveguide are evanescent. An important consequence is the characteristic impedance of the waveguide $Z_0$ will be imaginary in this case. \\
Substituting our expression for $k$ into eqs.~(\ref{eq:incident}) results in
\begin{gather}
    v_i(d) = a \times v_{in}H^{(2)}_0(-id/\delta_p)
    \label{eq:incidentplasma}
\end{gather}
where $\delta_p$ is the plasma penetration depth given in eq.~(\ref{eq:plasmapendepth}) in the main text.  For qubits weakly coupled to the waveguide mode,  reflections off qubit 2 back at qubit 1 will be small, and we ignore the reflected voltage and current wave.
\\
The voltage in the waveguide at the positions of qubit 1 and qubit 2 are then
\begin{gather}
    V(\delta_0) = v_{in}
    \label{eq:voltagestart}
    \\
    V(d_{12}-\delta_0) =a \times v_{in}H^{(2)}_0(-id_{12}/\delta_p)
\end{gather}
The small magnitude of the reflected voltage and current at $d=\delta_0$ means the input impedance of the waveguide is just $Z_0(\delta_0)$, where $Z_0$ is a function of $d$. Putting this together, we arrive at
\begin{gather}
    V_{d_{12}}/V_{0}= aH^{(2)}_0(-id_{12}/\delta_p)
    \\
    V_{0}/I_{0} = Z_0(\delta_0)
\end{gather}
We are now equipped to solve the circuit for $Z_{12}$. We define $Z_g=1/i\omega C_g,Z_{q1}=1/i\omega C_{q1},Z_{q2}=1/i\omega C_{q2}$, arriving at
\begin{equation}
    Z_{12}=\frac{Z_{q1}Z_{q2}Z_{0}(\delta_0)}{(Z_g+Z_{q2})(Z_g+Z_0(\delta_0))}\times aH^{(2)}_0(-id_{12}/\delta_p)
    \label{eq:transimpedance}
\end{equation}
where we have used $I_{q1}\approx V_{q1}/Z_{q1}$, $I_{q1}$ being the current applied through port 1. Since $Z_0$ is imaginary, we see $Z_{12}$ is also imaginary. For large coupling impedance $Z_g \gg Z_{q2},Z_0(\delta_0)$, eq.~(\ref{eq:transimpedance}) simplifies to
\begin{equation}
    Z_{12}=\frac{Z_{q1}Z_{q2}Z_{0}(\delta_0)}{Z_g^2}\times aH^{(2)}_0(-id_{12}/\delta_p)
    \label{eq:qubittransimpedance}
\end{equation}
We use the following expression for the coupling strength between qubits and the waveguide mode
\begin{equation}
    g = \frac{C_g'}{2}\sqrt{\frac{\omega_q v}{C_q C_r'(\delta_0)}} = \frac{C_g'}{2}\sqrt{\frac{\omega_q Z_0(\delta_0)}{C_q}}v
    \label{eq:qubitwaveguideg}
\end{equation}
where $v$ is the speed of light in the waveguide, $C_r'(\delta_0)$ is the capacitance per unit length of the waveguide mode at $\delta_0$, and $C_g'$ is the coupling capacitance per unit length between the qubit and the waveguide mode, where $C_g=C_g' \delta_0$. Substituting eqs.~(\ref{eq:qubittransimpedance}) \& (\ref{eq:qubitwaveguideg}) into eq.~(\ref{eq:JImpedanceFormula}), and using the relation $H_n^{(2)}(-jx)=2i K_n(x)/\pi$ where $K_n(x)$ is the modified Bessel function of the second kind, we find
\begin{equation}
J_{ij} =  2g^2\frac{\omega_q}{(v/\delta_0)^2}bK_0(d_{12}/\delta_p)
\end{equation}
where $b=1/K_0(\delta_0/\delta_p)$. Finally, we redefine $g\rightarrow g \times (K_0(\delta_0/\delta_p))^{1/2}$ to arrive at eq.~(\ref{eq:plasmacouplingstrength}) in the main text.
%%%% end of main
%%%% Bibliography included in main tex file

\bibliographystyle{apsrev4-1}

%%%%%%%%%%%%%%%%%%%%%%%%%%%%%%%%%%%%%%%%%%%%%


%\begin{thebibliography}{46}%
%\bibliography{References}

\begin{thebibliography}{35}%
\makeatletter
\providecommand \@ifxundefined [1]{%
 \@ifx{#1\undefined}
}%
\providecommand \@ifnum [1]{%
 \ifnum #1\expandafter \@firstoftwo
 \else \expandafter \@secondoftwo
 \fi
}%
\providecommand \@ifx [1]{%
 \ifx #1\expandafter \@firstoftwo
 \else \expandafter \@secondoftwo
 \fi
}%
\providecommand \natexlab [1]{#1}%
\providecommand \enquote  [1]{``#1''}%
\providecommand \bibnamefont  [1]{#1}%
\providecommand \bibfnamefont [1]{#1}%
\providecommand \citenamefont [1]{#1}%
\providecommand \href@noop [0]{\@secondoftwo}%
\providecommand \href [0]{\begingroup \@sanitize@url \@href}%
\providecommand \@href[1]{\@@startlink{#1}\@@href}%
\providecommand \@@href[1]{\endgroup#1\@@endlink}%
\providecommand \@sanitize@url [0]{\catcode `\\12\catcode `\$12\catcode
  `\&12\catcode `\#12\catcode `\^12\catcode `\_12\catcode `\%12\relax}%
\providecommand \@@startlink[1]{}%
\providecommand \@@endlink[0]{}%
\providecommand \url  [0]{\begingroup\@sanitize@url \@url }%
\providecommand \@url [1]{\endgroup\@href {#1}{\urlprefix }}%
\providecommand \urlprefix  [0]{URL }%
\providecommand \Eprint [0]{\href }%
\providecommand \doibase [0]{http://dx.doi.org/}%
\providecommand \selectlanguage [0]{\@gobble}%
\providecommand \bibinfo  [0]{\@secondoftwo}%
\providecommand \bibfield  [0]{\@secondoftwo}%
\providecommand \translation [1]{[#1]}%
\providecommand \BibitemOpen [0]{}%
\providecommand \bibitemStop [0]{}%
\providecommand \bibitemNoStop [0]{.\EOS\space}%
\providecommand \EOS [0]{\spacefactor3000\relax}%
\providecommand \BibitemShut  [1]{\csname bibitem#1\endcsname}%
\let\auto@bib@innerbib\@empty
%</preamble>
\bibitem [{\citenamefont {Sheldon}\ \emph {et~al.}(2016)\citenamefont
  {Sheldon}, \citenamefont {Bishop}, \citenamefont {Magesan}, \citenamefont
  {Filipp}, \citenamefont {Chow},\ and\ \citenamefont
  {Gambetta}}]{sheldon2016characterizing}%
  \BibitemOpen
  \bibfield  {author} {\bibinfo {author} {\bibfnamefont {S.}~\bibnamefont
  {Sheldon}}, \bibinfo {author} {\bibfnamefont {L.~S.}\ \bibnamefont {Bishop}},
  \bibinfo {author} {\bibfnamefont {E.}~\bibnamefont {Magesan}}, \bibinfo
  {author} {\bibfnamefont {S.}~\bibnamefont {Filipp}}, \bibinfo {author}
  {\bibfnamefont {J.~M.}\ \bibnamefont {Chow}}, \ and\ \bibinfo {author}
  {\bibfnamefont {J.~M.}\ \bibnamefont {Gambetta}},\ }\href@noop {} {\bibfield
  {journal} {\bibinfo  {journal} {Physical Review A}\ }\textbf {\bibinfo
  {volume} {93}},\ \bibinfo {pages} {012301} (\bibinfo {year}
  {2016})}\BibitemShut {NoStop}%
\bibitem [{\citenamefont {Barends}\ \emph {et~al.}(2019)\citenamefont
  {Barends}, \citenamefont {Quintana}, \citenamefont {Petukhov}, \citenamefont
  {Chen}, \citenamefont {Kafri}, \citenamefont {Kechedzhi}, \citenamefont
  {Collins}, \citenamefont {Naaman}, \citenamefont {Boixo}, \citenamefont
  {Arute} \emph {et~al.}}]{barends2019diabatic}%
  \BibitemOpen
  \bibfield  {author} {\bibinfo {author} {\bibfnamefont {R.}~\bibnamefont
  {Barends}}, \bibinfo {author} {\bibfnamefont {C.}~\bibnamefont {Quintana}},
  \bibinfo {author} {\bibfnamefont {A.}~\bibnamefont {Petukhov}}, \bibinfo
  {author} {\bibfnamefont {Y.}~\bibnamefont {Chen}}, \bibinfo {author}
  {\bibfnamefont {D.}~\bibnamefont {Kafri}}, \bibinfo {author} {\bibfnamefont
  {K.}~\bibnamefont {Kechedzhi}}, \bibinfo {author} {\bibfnamefont
  {R.}~\bibnamefont {Collins}}, \bibinfo {author} {\bibfnamefont
  {O.}~\bibnamefont {Naaman}}, \bibinfo {author} {\bibfnamefont
  {S.}~\bibnamefont {Boixo}}, \bibinfo {author} {\bibfnamefont
  {F.}~\bibnamefont {Arute}},  \emph {et~al.},\ }\href@noop {} {\bibfield
  {journal} {\bibinfo  {journal} {Physical Review Letters}\ }\textbf {\bibinfo
  {volume} {123}},\ \bibinfo {pages} {210501} (\bibinfo {year}
  {2019})}\BibitemShut {NoStop}%
\bibitem [{\citenamefont {Heinsoo}\ \emph {et~al.}(2018)\citenamefont
  {Heinsoo}, \citenamefont {Andersen}, \citenamefont {Remm}, \citenamefont
  {Krinner}, \citenamefont {Walter}, \citenamefont {Salath{\'e}}, \citenamefont
  {Gasparinetti}, \citenamefont {Besse}, \citenamefont {Poto{\v{c}}nik},
  \citenamefont {Wallraff} \emph {et~al.}}]{heinsoo2018rapid}%
  \BibitemOpen
  \bibfield  {author} {\bibinfo {author} {\bibfnamefont {J.}~\bibnamefont
  {Heinsoo}}, \bibinfo {author} {\bibfnamefont {C.~K.}\ \bibnamefont
  {Andersen}}, \bibinfo {author} {\bibfnamefont {A.}~\bibnamefont {Remm}},
  \bibinfo {author} {\bibfnamefont {S.}~\bibnamefont {Krinner}}, \bibinfo
  {author} {\bibfnamefont {T.}~\bibnamefont {Walter}}, \bibinfo {author}
  {\bibfnamefont {Y.}~\bibnamefont {Salath{\'e}}}, \bibinfo {author}
  {\bibfnamefont {S.}~\bibnamefont {Gasparinetti}}, \bibinfo {author}
  {\bibfnamefont {J.-C.}\ \bibnamefont {Besse}}, \bibinfo {author}
  {\bibfnamefont {A.}~\bibnamefont {Poto{\v{c}}nik}}, \bibinfo {author}
  {\bibfnamefont {A.}~\bibnamefont {Wallraff}},  \emph {et~al.},\ }\href@noop
  {} {\bibfield  {journal} {\bibinfo  {journal} {Physical Review Applied}\
  }\textbf {\bibinfo {volume} {10}},\ \bibinfo {pages} {034040} (\bibinfo
  {year} {2018})}\BibitemShut {NoStop}%
\bibitem [{\citenamefont {Reed}\ \emph {et~al.}(2010)\citenamefont {Reed},
  \citenamefont {Johnson}, \citenamefont {Houck}, \citenamefont {DiCarlo},
  \citenamefont {Chow}, \citenamefont {Schuster}, \citenamefont {Frunzio},\
  and\ \citenamefont {Schoelkopf}}]{reed2010fast}%
  \BibitemOpen
  \bibfield  {author} {\bibinfo {author} {\bibfnamefont {M.~D.}\ \bibnamefont
  {Reed}}, \bibinfo {author} {\bibfnamefont {B.~R.}\ \bibnamefont {Johnson}},
  \bibinfo {author} {\bibfnamefont {A.~A.}\ \bibnamefont {Houck}}, \bibinfo
  {author} {\bibfnamefont {L.}~\bibnamefont {DiCarlo}}, \bibinfo {author}
  {\bibfnamefont {J.~M.}\ \bibnamefont {Chow}}, \bibinfo {author}
  {\bibfnamefont {D.~I.}\ \bibnamefont {Schuster}}, \bibinfo {author}
  {\bibfnamefont {L.}~\bibnamefont {Frunzio}}, \ and\ \bibinfo {author}
  {\bibfnamefont {R.~J.}\ \bibnamefont {Schoelkopf}},\ }\href@noop {}
  {\bibfield  {journal} {\bibinfo  {journal} {Applied Physics Letters}\
  }\textbf {\bibinfo {volume} {96}},\ \bibinfo {pages} {203110} (\bibinfo
  {year} {2010})}\BibitemShut {NoStop}%
\bibitem [{\citenamefont {Houck}\ \emph {et~al.}(2008)\citenamefont {Houck},
  \citenamefont {Schreier}, \citenamefont {Johnson}, \citenamefont {Chow},
  \citenamefont {Koch}, \citenamefont {Gambetta}, \citenamefont {Schuster},
  \citenamefont {Frunzio}, \citenamefont {Devoret}, \citenamefont {Girvin}
  \emph {et~al.}}]{houck2008controlling}%
  \BibitemOpen
  \bibfield  {author} {\bibinfo {author} {\bibfnamefont {A.~A.}\ \bibnamefont
  {Houck}}, \bibinfo {author} {\bibfnamefont {J.~A.}\ \bibnamefont {Schreier}},
  \bibinfo {author} {\bibfnamefont {B.~R.}\ \bibnamefont {Johnson}}, \bibinfo
  {author} {\bibfnamefont {J.~M.}\ \bibnamefont {Chow}}, \bibinfo {author}
  {\bibfnamefont {J.}~\bibnamefont {Koch}}, \bibinfo {author} {\bibfnamefont
  {J.~M.}\ \bibnamefont {Gambetta}}, \bibinfo {author} {\bibfnamefont {D.~I.}\
  \bibnamefont {Schuster}}, \bibinfo {author} {\bibfnamefont {L.}~\bibnamefont
  {Frunzio}}, \bibinfo {author} {\bibfnamefont {M.~H.}\ \bibnamefont
  {Devoret}}, \bibinfo {author} {\bibfnamefont {S.~M.}\ \bibnamefont {Girvin}},
   \emph {et~al.},\ }\href@noop {} {\bibfield  {journal} {\bibinfo  {journal}
  {Physical review letters}\ }\textbf {\bibinfo {volume} {101}},\ \bibinfo
  {pages} {080502} (\bibinfo {year} {2008})}\BibitemShut {NoStop}%
\bibitem [{\citenamefont {McConkey}\ \emph {et~al.}(2018)\citenamefont
  {McConkey}, \citenamefont {B{\'e}janin}, \citenamefont {Earnest},
  \citenamefont {McRae}, \citenamefont {Pagel}, \citenamefont {Rinehart},\ and\
  \citenamefont {Mariantoni}}]{mcconkey2018mitigating}%
  \BibitemOpen
  \bibfield  {author} {\bibinfo {author} {\bibfnamefont {T.~G.}\ \bibnamefont
  {McConkey}}, \bibinfo {author} {\bibfnamefont {J.~H.}\ \bibnamefont
  {B{\'e}janin}}, \bibinfo {author} {\bibfnamefont {C.~T.}\ \bibnamefont
  {Earnest}}, \bibinfo {author} {\bibfnamefont {C.~R.~H.}\ \bibnamefont
  {McRae}}, \bibinfo {author} {\bibfnamefont {Z.}~\bibnamefont {Pagel}},
  \bibinfo {author} {\bibfnamefont {J.~R.}\ \bibnamefont {Rinehart}}, \ and\
  \bibinfo {author} {\bibfnamefont {M.}~\bibnamefont {Mariantoni}},\
  }\href@noop {} {\bibfield  {journal} {\bibinfo  {journal} {Quantum Science
  and Technology}\ }\textbf {\bibinfo {volume} {3}},\ \bibinfo {pages} {034004}
  (\bibinfo {year} {2018})}\BibitemShut {NoStop}%
\bibitem [{\citenamefont {Filipp}\ \emph {et~al.}(2011)\citenamefont {Filipp},
  \citenamefont {G{\"o}ppl}, \citenamefont {Fink}, \citenamefont {Baur},
  \citenamefont {Bianchetti}, \citenamefont {Steffen},\ and\ \citenamefont
  {Wallraff}}]{filipp2011multimode}%
  \BibitemOpen
  \bibfield  {author} {\bibinfo {author} {\bibfnamefont {S.}~\bibnamefont
  {Filipp}}, \bibinfo {author} {\bibfnamefont {M.}~\bibnamefont {G{\"o}ppl}},
  \bibinfo {author} {\bibfnamefont {J.~M.}\ \bibnamefont {Fink}}, \bibinfo
  {author} {\bibfnamefont {M.}~\bibnamefont {Baur}}, \bibinfo {author}
  {\bibfnamefont {R.}~\bibnamefont {Bianchetti}}, \bibinfo {author}
  {\bibfnamefont {L.}~\bibnamefont {Steffen}}, \ and\ \bibinfo {author}
  {\bibfnamefont {A.}~\bibnamefont {Wallraff}},\ }\href@noop {} {\bibfield
  {journal} {\bibinfo  {journal} {Physical Review A}\ }\textbf {\bibinfo
  {volume} {83}},\ \bibinfo {pages} {063827} (\bibinfo {year}
  {2011})}\BibitemShut {NoStop}%
\bibitem [{\citenamefont {Paik}\ \emph {et~al.}(2011)\citenamefont {Paik},
  \citenamefont {Schuster}, \citenamefont {Bishop}, \citenamefont {Kirchmair},
  \citenamefont {Catelani}, \citenamefont {Sears}, \citenamefont {Johnson},
  \citenamefont {Reagor}, \citenamefont {Frunzio}, \citenamefont {Glazman}
  \emph {et~al.}}]{paik2011observation}%
  \BibitemOpen
  \bibfield  {author} {\bibinfo {author} {\bibfnamefont {H.}~\bibnamefont
  {Paik}}, \bibinfo {author} {\bibfnamefont {D.~I.}\ \bibnamefont {Schuster}},
  \bibinfo {author} {\bibfnamefont {L.~S.}\ \bibnamefont {Bishop}}, \bibinfo
  {author} {\bibfnamefont {G.}~\bibnamefont {Kirchmair}}, \bibinfo {author}
  {\bibfnamefont {G.}~\bibnamefont {Catelani}}, \bibinfo {author}
  {\bibfnamefont {A.~P.}\ \bibnamefont {Sears}}, \bibinfo {author}
  {\bibfnamefont {B.}~\bibnamefont {Johnson}}, \bibinfo {author} {\bibfnamefont
  {M.~J.}\ \bibnamefont {Reagor}}, \bibinfo {author} {\bibfnamefont
  {L.}~\bibnamefont {Frunzio}}, \bibinfo {author} {\bibfnamefont {L.~I.}\
  \bibnamefont {Glazman}},  \emph {et~al.},\ }\href@noop {} {\bibfield
  {journal} {\bibinfo  {journal} {Physical Review Letters}\ }\textbf {\bibinfo
  {volume} {107}},\ \bibinfo {pages} {240501} (\bibinfo {year}
  {2011})}\BibitemShut {NoStop}%
\bibitem [{\citenamefont {Bronn}\ \emph {et~al.}(2018)\citenamefont {Bronn},
  \citenamefont {Adiga}, \citenamefont {Olivadese}, \citenamefont {Wu},
  \citenamefont {Chow},\ and\ \citenamefont {Pappas}}]{bronn2018high}%
  \BibitemOpen
  \bibfield  {author} {\bibinfo {author} {\bibfnamefont {N.~T.}\ \bibnamefont
  {Bronn}}, \bibinfo {author} {\bibfnamefont {V.~P.}\ \bibnamefont {Adiga}},
  \bibinfo {author} {\bibfnamefont {S.~B.}\ \bibnamefont {Olivadese}}, \bibinfo
  {author} {\bibfnamefont {X.}~\bibnamefont {Wu}}, \bibinfo {author}
  {\bibfnamefont {J.~M.}\ \bibnamefont {Chow}}, \ and\ \bibinfo {author}
  {\bibfnamefont {D.~P.}\ \bibnamefont {Pappas}},\ }\href@noop {} {\bibfield
  {journal} {\bibinfo  {journal} {Quantum science and technology}\ }\textbf
  {\bibinfo {volume} {3}},\ \bibinfo {pages} {024007} (\bibinfo {year}
  {2018})}\BibitemShut {NoStop}%
\bibitem [{\citenamefont {Wenner}\ \emph {et~al.}(2011)\citenamefont {Wenner},
  \citenamefont {Neeley}, \citenamefont {Bialczak}, \citenamefont {Lenander},
  \citenamefont {Lucero}, \citenamefont {O'Connell}, \citenamefont {Sank},
  \citenamefont {Wang}, \citenamefont {Weides}, \citenamefont {Cleland} \emph
  {et~al.}}]{wenner2011wirebond}%
  \BibitemOpen
  \bibfield  {author} {\bibinfo {author} {\bibfnamefont {J.}~\bibnamefont
  {Wenner}}, \bibinfo {author} {\bibfnamefont {M.}~\bibnamefont {Neeley}},
  \bibinfo {author} {\bibfnamefont {R.~C.}\ \bibnamefont {Bialczak}}, \bibinfo
  {author} {\bibfnamefont {M.}~\bibnamefont {Lenander}}, \bibinfo {author}
  {\bibfnamefont {E.}~\bibnamefont {Lucero}}, \bibinfo {author} {\bibfnamefont
  {A.~D.}\ \bibnamefont {O'Connell}}, \bibinfo {author} {\bibfnamefont
  {D.}~\bibnamefont {Sank}}, \bibinfo {author} {\bibfnamefont {H.}~\bibnamefont
  {Wang}}, \bibinfo {author} {\bibfnamefont {M.}~\bibnamefont {Weides}},
  \bibinfo {author} {\bibfnamefont {A.~N.}\ \bibnamefont {Cleland}},  \emph
  {et~al.},\ }\href@noop {} {\bibfield  {journal} {\bibinfo  {journal}
  {Superconductor Science and Technology}\ }\textbf {\bibinfo {volume} {24}},\
  \bibinfo {pages} {065001} (\bibinfo {year} {2011})}\BibitemShut {NoStop}%
\bibitem [{\citenamefont {Gambetta}\ \emph {et~al.}(2017)\citenamefont
  {Gambetta}, \citenamefont {Chow},\ and\ \citenamefont
  {Steffen}}]{gambetta2017building}%
  \BibitemOpen
  \bibfield  {author} {\bibinfo {author} {\bibfnamefont {J.~M.}\ \bibnamefont
  {Gambetta}}, \bibinfo {author} {\bibfnamefont {J.~M.}\ \bibnamefont {Chow}},
  \ and\ \bibinfo {author} {\bibfnamefont {M.}~\bibnamefont {Steffen}},\
  }\href@noop {} {\bibfield  {journal} {\bibinfo  {journal} {npj Quantum
  Information}\ }\textbf {\bibinfo {volume} {3}},\ \bibinfo {pages} {2}
  (\bibinfo {year} {2017})}\BibitemShut {NoStop}%
\bibitem [{\citenamefont {Vahidpour}\ \emph {et~al.}(2017)\citenamefont
  {Vahidpour}, \citenamefont {O'Brien}, \citenamefont {Whyland}, \citenamefont
  {Angeles}, \citenamefont {Marshall}, \citenamefont {Scarabelli},
  \citenamefont {Crossman}, \citenamefont {Yadav}, \citenamefont {Mohan},
  \citenamefont {Bui} \emph {et~al.}}]{vahidpour2017superconducting}%
  \BibitemOpen
  \bibfield  {author} {\bibinfo {author} {\bibfnamefont {M.}~\bibnamefont
  {Vahidpour}}, \bibinfo {author} {\bibfnamefont {W.}~\bibnamefont {O'Brien}},
  \bibinfo {author} {\bibfnamefont {J.~T.}\ \bibnamefont {Whyland}}, \bibinfo
  {author} {\bibfnamefont {J.}~\bibnamefont {Angeles}}, \bibinfo {author}
  {\bibfnamefont {J.}~\bibnamefont {Marshall}}, \bibinfo {author}
  {\bibfnamefont {D.}~\bibnamefont {Scarabelli}}, \bibinfo {author}
  {\bibfnamefont {G.}~\bibnamefont {Crossman}}, \bibinfo {author}
  {\bibfnamefont {K.}~\bibnamefont {Yadav}}, \bibinfo {author} {\bibfnamefont
  {Y.}~\bibnamefont {Mohan}}, \bibinfo {author} {\bibfnamefont
  {C.}~\bibnamefont {Bui}},  \emph {et~al.},\ }\href@noop {} {\bibfield
  {journal} {\bibinfo  {journal} {arXiv preprint arXiv:1708.02226}\ } (\bibinfo
  {year} {2017})}\BibitemShut {NoStop}%
\bibitem [{\citenamefont {Yost}\ \emph {et~al.}(2019)\citenamefont {Yost},
  \citenamefont {Schwartz}, \citenamefont {Mallek}, \citenamefont {Rosenberg},
  \citenamefont {Stull}, \citenamefont {Yoder}, \citenamefont {Calusine},
  \citenamefont {Cook}, \citenamefont {Das}, \citenamefont {Day} \emph
  {et~al.}}]{yost2019solid}%
  \BibitemOpen
  \bibfield  {author} {\bibinfo {author} {\bibfnamefont {D.-R.~W.}\
  \bibnamefont {Yost}}, \bibinfo {author} {\bibfnamefont {M.~E.}\ \bibnamefont
  {Schwartz}}, \bibinfo {author} {\bibfnamefont {J.}~\bibnamefont {Mallek}},
  \bibinfo {author} {\bibfnamefont {D.}~\bibnamefont {Rosenberg}}, \bibinfo
  {author} {\bibfnamefont {C.}~\bibnamefont {Stull}}, \bibinfo {author}
  {\bibfnamefont {J.~L.}\ \bibnamefont {Yoder}}, \bibinfo {author}
  {\bibfnamefont {G.}~\bibnamefont {Calusine}}, \bibinfo {author}
  {\bibfnamefont {M.}~\bibnamefont {Cook}}, \bibinfo {author} {\bibfnamefont
  {R.}~\bibnamefont {Das}}, \bibinfo {author} {\bibfnamefont {A.~L.}\
  \bibnamefont {Day}},  \emph {et~al.},\ }\href@noop {} {\bibfield  {journal}
  {\bibinfo  {journal} {arXiv preprint arXiv:1912.10942}\ } (\bibinfo {year}
  {2019})}\BibitemShut {NoStop}%
\bibitem [{\citenamefont {Nicorovici}\ \emph {et~al.}(1995)\citenamefont
  {Nicorovici}, \citenamefont {McPhedran},\ and\ \citenamefont
  {Botten}}]{nicorovici1995photonic}%
  \BibitemOpen
  \bibfield  {author} {\bibinfo {author} {\bibfnamefont {N.~A.}\ \bibnamefont
  {Nicorovici}}, \bibinfo {author} {\bibfnamefont {R.~C.}\ \bibnamefont
  {McPhedran}}, \ and\ \bibinfo {author} {\bibfnamefont {L.~C.}\ \bibnamefont
  {Botten}},\ }\href@noop {} {\bibfield  {journal} {\bibinfo  {journal}
  {Physical Review E}\ }\textbf {\bibinfo {volume} {52}},\ \bibinfo {pages}
  {1135} (\bibinfo {year} {1995})}\BibitemShut {NoStop}%
\bibitem [{\citenamefont {Smith}\ \emph {et~al.}(1994)\citenamefont {Smith},
  \citenamefont {Schultz}, \citenamefont {Kroll}, \citenamefont {Sigalas},
  \citenamefont {Ho},\ and\ \citenamefont {Soukoulis}}]{smith1994experimental}%
  \BibitemOpen
  \bibfield  {author} {\bibinfo {author} {\bibfnamefont {D.~R.}\ \bibnamefont
  {Smith}}, \bibinfo {author} {\bibfnamefont {S.}~\bibnamefont {Schultz}},
  \bibinfo {author} {\bibfnamefont {N.}~\bibnamefont {Kroll}}, \bibinfo
  {author} {\bibfnamefont {M.}~\bibnamefont {Sigalas}}, \bibinfo {author}
  {\bibfnamefont {K.~M.}\ \bibnamefont {Ho}}, \ and\ \bibinfo {author}
  {\bibfnamefont {C.~M.}\ \bibnamefont {Soukoulis}},\ }\href@noop {} {\bibfield
   {journal} {\bibinfo  {journal} {Applied Physics Letters}\ }\textbf {\bibinfo
  {volume} {65}},\ \bibinfo {pages} {645} (\bibinfo {year} {1994})}\BibitemShut
  {NoStop}%
\bibitem [{\citenamefont {Preskill}(2018)}]{preskill2018quantum}%
  \BibitemOpen
  \bibfield  {author} {\bibinfo {author} {\bibfnamefont {J.}~\bibnamefont
  {Preskill}},\ }\href@noop {} {\bibfield  {journal} {\bibinfo  {journal}
  {Quantum}\ }\textbf {\bibinfo {volume} {2}},\ \bibinfo {pages} {79} (\bibinfo
  {year} {2018})}\BibitemShut {NoStop}%
\bibitem [{\citenamefont {Pendry}\ \emph {et~al.}(1996)\citenamefont {Pendry},
  \citenamefont {Holden}, \citenamefont {Stewart},\ and\ \citenamefont
  {Youngs}}]{pendry1996extremely}%
  \BibitemOpen
  \bibfield  {author} {\bibinfo {author} {\bibfnamefont {J.~B.}\ \bibnamefont
  {Pendry}}, \bibinfo {author} {\bibfnamefont {A.~J.}\ \bibnamefont {Holden}},
  \bibinfo {author} {\bibfnamefont {W.~J.}\ \bibnamefont {Stewart}}, \ and\
  \bibinfo {author} {\bibfnamefont {I.}~\bibnamefont {Youngs}},\ }\href@noop {}
  {\bibfield  {journal} {\bibinfo  {journal} {Physical review letters}\
  }\textbf {\bibinfo {volume} {76}},\ \bibinfo {pages} {4773} (\bibinfo {year}
  {1996})}\BibitemShut {NoStop}%
\bibitem [{\citenamefont {Belov}\ \emph {et~al.}(2003)\citenamefont {Belov},
  \citenamefont {Marques}, \citenamefont {Maslovski}, \citenamefont {Nefedov},
  \citenamefont {Silveirinha}, \citenamefont {Simovski},\ and\ \citenamefont
  {Tretyakov}}]{belov2003strong}%
  \BibitemOpen
  \bibfield  {author} {\bibinfo {author} {\bibfnamefont {P.~A.}\ \bibnamefont
  {Belov}}, \bibinfo {author} {\bibfnamefont {R.}~\bibnamefont {Marques}},
  \bibinfo {author} {\bibfnamefont {S.~I.}\ \bibnamefont {Maslovski}}, \bibinfo
  {author} {\bibfnamefont {I.~S.}\ \bibnamefont {Nefedov}}, \bibinfo {author}
  {\bibfnamefont {M.}~\bibnamefont {Silveirinha}}, \bibinfo {author}
  {\bibfnamefont {C.~R.}\ \bibnamefont {Simovski}}, \ and\ \bibinfo {author}
  {\bibfnamefont {S.~A.}\ \bibnamefont {Tretyakov}},\ }\href@noop {} {\bibfield
   {journal} {\bibinfo  {journal} {Physical Review B}\ }\textbf {\bibinfo
  {volume} {67}},\ \bibinfo {pages} {113103} (\bibinfo {year}
  {2003})}\BibitemShut {NoStop}%
\bibitem [{\citenamefont {Belov}\ \emph {et~al.}(2002)\citenamefont {Belov},
  \citenamefont {Tretyakov},\ and\ \citenamefont
  {Viitanen}}]{belov2002dispersion}%
  \BibitemOpen
  \bibfield  {author} {\bibinfo {author} {\bibfnamefont {P.~A.}\ \bibnamefont
  {Belov}}, \bibinfo {author} {\bibfnamefont {S.~A.}\ \bibnamefont
  {Tretyakov}}, \ and\ \bibinfo {author} {\bibfnamefont {A.~J.}\ \bibnamefont
  {Viitanen}},\ }\href@noop {} {\bibfield  {journal} {\bibinfo  {journal}
  {Journal of electromagnetic waves and applications}\ }\textbf {\bibinfo
  {volume} {16}},\ \bibinfo {pages} {1153} (\bibinfo {year}
  {2002})}\BibitemShut {NoStop}%
\bibitem [{\citenamefont {Krynkin}\ and\ \citenamefont
  {McIver}(2009)}]{krynkin2009approximations}%
  \BibitemOpen
  \bibfield  {author} {\bibinfo {author} {\bibfnamefont {A.}~\bibnamefont
  {Krynkin}}\ and\ \bibinfo {author} {\bibfnamefont {P.}~\bibnamefont
  {McIver}},\ }\href@noop {} {\bibfield  {journal} {\bibinfo  {journal} {Waves
  in Random and Complex Media}\ }\textbf {\bibinfo {volume} {19}},\ \bibinfo
  {pages} {347} (\bibinfo {year} {2009})}\BibitemShut {NoStop}%
\bibitem [{\citenamefont {Murray}\ and\ \citenamefont
  {Abraham}(2016)}]{murray2016predicting}%
  \BibitemOpen
  \bibfield  {author} {\bibinfo {author} {\bibfnamefont {C.~E.}\ \bibnamefont
  {Murray}}\ and\ \bibinfo {author} {\bibfnamefont {D.~W.}\ \bibnamefont
  {Abraham}},\ }\href@noop {} {\bibfield  {journal} {\bibinfo  {journal}
  {Applied Physics Letters}\ }\textbf {\bibinfo {volume} {108}},\ \bibinfo
  {pages} {084101} (\bibinfo {year} {2016})}\BibitemShut {NoStop}%
\bibitem [{\citenamefont {Pendry}\ \emph {et~al.}(1998)\citenamefont {Pendry},
  \citenamefont {Holden}, \citenamefont {Robbins},\ and\ \citenamefont
  {Stewart}}]{pendry1998low}%
  \BibitemOpen
  \bibfield  {author} {\bibinfo {author} {\bibfnamefont {J.~B.}\ \bibnamefont
  {Pendry}}, \bibinfo {author} {\bibfnamefont {A.~J.}\ \bibnamefont {Holden}},
  \bibinfo {author} {\bibfnamefont {D.~J.}\ \bibnamefont {Robbins}}, \ and\
  \bibinfo {author} {\bibfnamefont {W.~J.}\ \bibnamefont {Stewart}},\
  }\href@noop {} {\bibfield  {journal} {\bibinfo  {journal} {Journal of
  Physics: Condensed Matter}\ }\textbf {\bibinfo {volume} {10}},\ \bibinfo
  {pages} {4785} (\bibinfo {year} {1998})}\BibitemShut {NoStop}%
\bibitem [{\citenamefont {Remski}(2000)}]{remski2000analysis}%
  \BibitemOpen
  \bibfield  {author} {\bibinfo {author} {\bibfnamefont {R.}~\bibnamefont
  {Remski}},\ }\href@noop {} {\bibfield  {journal} {\bibinfo  {journal}
  {Microwave Journal}\ }\textbf {\bibinfo {volume} {43}},\ \bibinfo {pages}
  {190} (\bibinfo {year} {2000})}\BibitemShut {NoStop}%
\bibitem [{\citenamefont {Hartmann}\ \emph {et~al.}(2006)\citenamefont
  {Hartmann}, \citenamefont {Brandao},\ and\ \citenamefont
  {Plenio}}]{hartmann2006strongly}%
  \BibitemOpen
  \bibfield  {author} {\bibinfo {author} {\bibfnamefont {M.~J.}\ \bibnamefont
  {Hartmann}}, \bibinfo {author} {\bibfnamefont {F.~G.}\ \bibnamefont
  {Brandao}}, \ and\ \bibinfo {author} {\bibfnamefont {M.~B.}\ \bibnamefont
  {Plenio}},\ }\href@noop {} {\bibfield  {journal} {\bibinfo  {journal} {Nature
  Physics}\ }\textbf {\bibinfo {volume} {2}},\ \bibinfo {pages} {849} (\bibinfo
  {year} {2006})}\BibitemShut {NoStop}%
\bibitem [{\citenamefont {Nagle}\ \emph {et~al.}(1967)\citenamefont {Nagle},
  \citenamefont {Knapp},\ and\ \citenamefont {Knapp}}]{nagle1967coupled}%
  \BibitemOpen
  \bibfield  {author} {\bibinfo {author} {\bibfnamefont {D.~E.}\ \bibnamefont
  {Nagle}}, \bibinfo {author} {\bibfnamefont {E.~A.}\ \bibnamefont {Knapp}}, \
  and\ \bibinfo {author} {\bibfnamefont {B.~C.}\ \bibnamefont {Knapp}},\
  }\href@noop {} {\bibfield  {journal} {\bibinfo  {journal} {Review of
  Scientific Instruments}\ }\textbf {\bibinfo {volume} {38}},\ \bibinfo {pages}
  {1583} (\bibinfo {year} {1967})}\BibitemShut {NoStop}%
\bibitem [{\citenamefont {Wangler}(2008)}]{wangler2008rf}%
  \BibitemOpen
  \bibfield  {author} {\bibinfo {author} {\bibfnamefont {T.~P.}\ \bibnamefont
  {Wangler}},\ }\href@noop {} {\emph {\bibinfo {title} {RF Linear
  accelerators}}}\ (\bibinfo  {publisher} {John Wiley \& Sons},\ \bibinfo
  {year} {2008})\BibitemShut {NoStop}%
\bibitem [{\citenamefont {Marcuvitz}(1951)}]{marcuvitz1951waveguide}%
  \BibitemOpen
  \bibfield  {author} {\bibinfo {author} {\bibfnamefont {N.}~\bibnamefont
  {Marcuvitz}},\ }\href@noop {} {\emph {\bibinfo {title} {Waveguide
  handbook}}},\ \bibinfo {number} {21}\ (\bibinfo  {publisher} {Iet},\ \bibinfo
  {year} {1951})\BibitemShut {NoStop}%
\bibitem [{\citenamefont {Shi}\ \emph {et~al.}(2016)\citenamefont {Shi},
  \citenamefont {Wu}, \citenamefont {Gonz{\'a}lez-Tudela},\ and\ \citenamefont
  {Cirac}}]{shi2016bound}%
  \BibitemOpen
  \bibfield  {author} {\bibinfo {author} {\bibfnamefont {T.}~\bibnamefont
  {Shi}}, \bibinfo {author} {\bibfnamefont {Y.-H.}\ \bibnamefont {Wu}},
  \bibinfo {author} {\bibfnamefont {A.}~\bibnamefont {Gonz{\'a}lez-Tudela}}, \
  and\ \bibinfo {author} {\bibfnamefont {J.~I.}\ \bibnamefont {Cirac}},\
  }\href@noop {} {\bibfield  {journal} {\bibinfo  {journal} {Physical Review
  X}\ }\textbf {\bibinfo {volume} {6}},\ \bibinfo {pages} {021027} (\bibinfo
  {year} {2016})}\BibitemShut {NoStop}%
\bibitem [{\citenamefont {Douglas}\ \emph {et~al.}(2015)\citenamefont
  {Douglas}, \citenamefont {Habibian}, \citenamefont {Hung}, \citenamefont
  {Gorshkov}, \citenamefont {Kimble},\ and\ \citenamefont
  {Chang}}]{douglas2015quantum}%
  \BibitemOpen
  \bibfield  {author} {\bibinfo {author} {\bibfnamefont {J.~S.}\ \bibnamefont
  {Douglas}}, \bibinfo {author} {\bibfnamefont {H.}~\bibnamefont {Habibian}},
  \bibinfo {author} {\bibfnamefont {C.-L.}\ \bibnamefont {Hung}}, \bibinfo
  {author} {\bibfnamefont {A.~V.}\ \bibnamefont {Gorshkov}}, \bibinfo {author}
  {\bibfnamefont {H.~J.}\ \bibnamefont {Kimble}}, \ and\ \bibinfo {author}
  {\bibfnamefont {D.~E.}\ \bibnamefont {Chang}},\ }\href@noop {} {\bibfield
  {journal} {\bibinfo  {journal} {Nature Photonics}\ }\textbf {\bibinfo
  {volume} {9}},\ \bibinfo {pages} {326} (\bibinfo {year} {2015})}\BibitemShut
  {NoStop}%
\bibitem [{\citenamefont {Gonz{\'a}lez-Tudela}\ \emph
  {et~al.}(2015)\citenamefont {Gonz{\'a}lez-Tudela}, \citenamefont {Hung},
  \citenamefont {Chang}, \citenamefont {Cirac},\ and\ \citenamefont
  {Kimble}}]{gonzalez2015subwavelength}%
  \BibitemOpen
  \bibfield  {author} {\bibinfo {author} {\bibfnamefont {A.}~\bibnamefont
  {Gonz{\'a}lez-Tudela}}, \bibinfo {author} {\bibfnamefont {C.-L.}\
  \bibnamefont {Hung}}, \bibinfo {author} {\bibfnamefont {D.~E.}\ \bibnamefont
  {Chang}}, \bibinfo {author} {\bibfnamefont {J.~I.}\ \bibnamefont {Cirac}}, \
  and\ \bibinfo {author} {\bibfnamefont {H.~J.}\ \bibnamefont {Kimble}},\
  }\href@noop {} {\bibfield  {journal} {\bibinfo  {journal} {Nature Photonics}\
  }\textbf {\bibinfo {volume} {9}},\ \bibinfo {pages} {320} (\bibinfo {year}
  {2015})}\BibitemShut {NoStop}%
\bibitem [{\citenamefont {Rahamim}\ \emph {et~al.}(2017)\citenamefont
  {Rahamim}, \citenamefont {Behrle}, \citenamefont {Peterer}, \citenamefont
  {Patterson}, \citenamefont {Spring}, \citenamefont {Tsunoda}, \citenamefont
  {Manenti}, \citenamefont {Tancredi},\ and\ \citenamefont
  {Leek}}]{rahamim2017double}%
  \BibitemOpen
  \bibfield  {author} {\bibinfo {author} {\bibfnamefont {J.}~\bibnamefont
  {Rahamim}}, \bibinfo {author} {\bibfnamefont {T.}~\bibnamefont {Behrle}},
  \bibinfo {author} {\bibfnamefont {M.~J.}\ \bibnamefont {Peterer}}, \bibinfo
  {author} {\bibfnamefont {A.}~\bibnamefont {Patterson}}, \bibinfo {author}
  {\bibfnamefont {P.~A.}\ \bibnamefont {Spring}}, \bibinfo {author}
  {\bibfnamefont {T.}~\bibnamefont {Tsunoda}}, \bibinfo {author} {\bibfnamefont
  {R.}~\bibnamefont {Manenti}}, \bibinfo {author} {\bibfnamefont
  {G.}~\bibnamefont {Tancredi}}, \ and\ \bibinfo {author} {\bibfnamefont
  {P.~J.}\ \bibnamefont {Leek}},\ }\href@noop {} {\bibfield  {journal}
  {\bibinfo  {journal} {Applied Physics Letters}\ }\textbf {\bibinfo {volume}
  {110}},\ \bibinfo {pages} {222602} (\bibinfo {year} {2017})}\BibitemShut
  {NoStop}%
\bibitem [{\citenamefont {Patterson}\ \emph {et~al.}(2019)\citenamefont
  {Patterson}, \citenamefont {Rahamim}, \citenamefont {Tsunoda}, \citenamefont
  {Spring}, \citenamefont {Jebari}, \citenamefont {Ratter}, \citenamefont
  {Mergenthaler}, \citenamefont {Tancredi}, \citenamefont {Vlastakis},
  \citenamefont {Esposito} \emph {et~al.}}]{patterson2019calibration}%
  \BibitemOpen
  \bibfield  {author} {\bibinfo {author} {\bibfnamefont {A.}~\bibnamefont
  {Patterson}}, \bibinfo {author} {\bibfnamefont {J.}~\bibnamefont {Rahamim}},
  \bibinfo {author} {\bibfnamefont {T.}~\bibnamefont {Tsunoda}}, \bibinfo
  {author} {\bibfnamefont {P.}~\bibnamefont {Spring}}, \bibinfo {author}
  {\bibfnamefont {S.}~\bibnamefont {Jebari}}, \bibinfo {author} {\bibfnamefont
  {K.}~\bibnamefont {Ratter}}, \bibinfo {author} {\bibfnamefont
  {M.}~\bibnamefont {Mergenthaler}}, \bibinfo {author} {\bibfnamefont
  {G.}~\bibnamefont {Tancredi}}, \bibinfo {author} {\bibfnamefont
  {B.}~\bibnamefont {Vlastakis}}, \bibinfo {author} {\bibfnamefont
  {M.}~\bibnamefont {Esposito}},  \emph {et~al.},\ }\href@noop {} {\bibfield
  {journal} {\bibinfo  {journal} {Physical Review Applied}\ }\textbf {\bibinfo
  {volume} {12}},\ \bibinfo {pages} {064013} (\bibinfo {year}
  {2019})}\BibitemShut {NoStop}%
\bibitem [{\citenamefont {Solgun}\ \emph {et~al.}(2019)\citenamefont {Solgun},
  \citenamefont {DiVincenzo},\ and\ \citenamefont
  {Gambetta}}]{solgun2019simple}%
  \BibitemOpen
  \bibfield  {author} {\bibinfo {author} {\bibfnamefont {F.}~\bibnamefont
  {Solgun}}, \bibinfo {author} {\bibfnamefont {D.~P.}\ \bibnamefont
  {DiVincenzo}}, \ and\ \bibinfo {author} {\bibfnamefont {J.~M.}\ \bibnamefont
  {Gambetta}},\ }\href@noop {} {\bibfield  {journal} {\bibinfo  {journal} {IEEE
  Transactions on Microwave Theory and Techniques}\ } (\bibinfo {year}
  {2019})}\BibitemShut {NoStop}%
\bibitem [{\citenamefont {Hayt}\ \emph {et~al.}(1978)\citenamefont {Hayt},
  \citenamefont {Kemmerly},\ and\ \citenamefont
  {Durbin}}]{hayt1978engineering}%
  \BibitemOpen
  \bibfield  {author} {\bibinfo {author} {\bibfnamefont {W.~H.}\ \bibnamefont
  {Hayt}}, \bibinfo {author} {\bibfnamefont {J.~E.}\ \bibnamefont {Kemmerly}},
  \ and\ \bibinfo {author} {\bibfnamefont {S.~M.}\ \bibnamefont {Durbin}},\
  }\href@noop {} {\emph {\bibinfo {title} {Engineering circuit analysis}}},\
  Vol.\ \bibinfo {volume} {214}\ (\bibinfo  {publisher} {McGraw-Hill New
  York},\ \bibinfo {year} {1978})\BibitemShut {NoStop}%
\bibitem [{\citenamefont {Losonczi}(1992)}]{losonczi1992eigenvalues}%
  \BibitemOpen
  \bibfield  {author} {\bibinfo {author} {\bibfnamefont {L.}~\bibnamefont
  {Losonczi}},\ }\href@noop {} {\bibfield  {journal} {\bibinfo  {journal} {Acta
  Mathematica Hungarica}\ }\textbf {\bibinfo {volume} {60}},\ \bibinfo {pages}
  {309} (\bibinfo {year} {1992})}\BibitemShut {NoStop}%
\end{thebibliography}
%\end{thebibliography}

\end{document}